\documentclass[fleqn,usenatbib]{mnras}

\usepackage{newtxtext,newtxmath}

\usepackage[T1]{fontenc}
\usepackage{ae,aecompl}

\usepackage{graphicx}	
\usepackage{hyperref}
\usepackage{amsmath}	

\def\iso#1#2{\mbox{${}^{#2}{\rm #1}$}}
\def\al2#1{\iso{Al}{2#1}}

\def\RSN{{\cal R}}

\def\pfrac#1#2{\left( \frac{#1}{#2} \right)}

\def\mVpeak{m_V^{\rm peak}}

\title[Rates and Detectability of Historical Supernovae]{Witnessing History: Rates and Detectability of Naked-Eye Milky-Way Supernovae}

\author[C.T. Murphey, J.W. Hogan , B.D. Fields, G. Narayan]{
C. Tanner Murphey$^{1,3,4}$,
Jacob W. Hogan$^{1,3,4}$,
Brian D. Fields$^{1,2,3,4}$,
Gautham Narayan$^{1,2,3,4}$
\\
$^{1}$Department of Astronomy, University of Illinois, Urbana, IL 61801 \\
$^{2}$Department of Physics, University of Illinois, Urbana, IL 61801 \\
$^{3}$Illinois Center for Advanced Studies of the Universe, University of Illinois, Urbana, IL 61801 \\
$^{4}$Center for AstroPhysical Surveys, National Center for Supercomputing Applications, Urbana, IL, 61801, USA \\
}

\date{Accepted XXX. Received YYY; in original form ZZZ}

\pubyear{2020}

\begin{document}
\label{firstpage}
\pagerange{\pageref{firstpage}--\pageref{lastpage}}
\maketitle

\begin{abstract}
The Milky Way hosts on average a few supernova explosions per century, yet in the past millennium
only five supernovae have been identified confidently in the historical
record.  This deficit of naked-eye supernovae
is at least partly due to
dust extinction in the Galactic plane.
We explore this effect quantitatively, developing a formalism for
the supernova probability distribution, accounting for dust and for
the observer's flux limit.
We then construct a fiducial axisymmetric model for the supernova and dust
densities, featuring an exponential dependence on galactocentric radius and height,
with core-collapse events in a thin disk and Type Ia events including a thick disk component. 
When no flux limit is applied, 
our model predicts supernovae are intrinsically concentrated in the Galactic plane,
with Type Ia events extending to higher
latitudes reflecting their thick disk component.
We then apply a flux limit and include dust effects, to predict the 
sky distribution of historical supernovae. 
We use well-observed supernovae as light-curve templates,
and introduce naked-eye discovery criteria.
The resulting sky distributions are strikingly inconsistent 
with the locations of confident historical supernovae, none of which
lie near our model's central peaks. Indeed, SN 1054  lies off the plane almost exactly in the anticenter, and SN 1181 is in the 2nd Galactic quadrant. We discuss possible explanations for these discrepancies.
We calculate the percentage of all supernovae bright enough for historical
discovery:
$\simeq 13\%$ of core-collapse and $\simeq 33\%$ of Type Ia events.
Using these and the confident historical supernovae,
we estimate the intrinsic 
Galactic supernova rates, finding general agreement with other methods.
Finally, we urge searches for supernovae in historical records from civilizations in the southern hemisphere.
\end{abstract}




\section{Introduction}
A Galactic supernova (SN) has not been convincingly observed 
by the naked eye since Kepler's SN 1604 \citep{kepler}.
This 416 year gap is disappointing, and persists despite a long and fruitful effort to identify supernovae (SNe) in the historical record \citep[e.g.,][]{Clark1977, Stephenson2002, guest_stars}. 
Over the four intervening centuries since Kepler's event,
modern Astronomy has arisen,
bringing advancements in telescopes and observational techniques that should have, if anything, made it more likely that Galactic SNe would be noticed. Rates and observability of Galactic SNe are clearly worthy of investigation.

The important study of \citet{adams}
anticipates the next Galactic SN, 
and uses historical data to
estimate Galactic rates for both core-collapse and thermonuclear or
Type Ia supernovae (hereafter CCSNe and SNIa, respectively). 
They conclude that the total rate of Galactic SNe corresponds to an average of one event per $\sim 22$ years (Table \ref{tab:sn_types}). 
Based on previously predicted rates, $\sim 18$ Galactic SNe should have occurred since Kepler's. 
Other methods give Galactic SN rates that can
vary by a factor $\sim 2$ \citep{Rozwadowska2021},
but even within these uncertainties it is extremely unlikely that no SN has occurred in the past 416 years.
Of course, SNe preferentially occur within 
the Milky Way thin disk where dust is also concentrated, and so extinction will play
an important role for distant events.
It is thus widely acknowledged that {\em historic supernovae}--i.e., naked-eye events discernible in past records--should represent only 
a fraction of Galactic events; this is
the subject of our study.

Observations of supernova remnants confirm that unseen events have
occurred since Kepler.  
The Cassiopeia A remnant
has a measured expansion age of about 325 years, corresponding to an explosion in $1681 \pm 19$ \citep{Fesen2006}.  It certainly was not widely observed; there have been claims that John Flamsteed saw it in 1680, but 
it is likely that he did not \citep{flamsteed}.
As far as we know, the most recent Galactic supernova remnant (SNR) claimed in the literature is G1.9+0.3 with an age of about $\sim 110-180 \ \rm yr$, which places the explosion in the latter half of the 19th century \citep{g1903_1, g1903_2, g1903_3}. 
Thus, at least two Galactic supernovae have
occurred since 1604.

To our knowledge, there are
no remnants that are dated to be younger than 
G1.9+0.3, suggesting that there have been no Galactic events in the 20th or 21st centuries. 
During this time,
technologies have been developed to detect wavelengths invisible human eye in which the Galaxy is partly or entirely transparent: infrared, radio, and gamma-ray. Moreover, core-collapse events would be detected via neutrino monitoring that has been ongoing for around 40 years
\citep{Antonioli2004,Ikeda2007,Novoseltsev2020,Scholberg2012,DUNE2020,SNEWS2020}, 
and now gravitational wave monitoring is provided by LIGO/VIRGO \citep{Abbott2020}.
Type Ia events likely would escape these searches but would be found
in gamma-ray burst monitors via their MeV line emission \citep{Wang2019}.
We can thus confidently look forward to witnessing the inevitable occurrence of the next Galactic supernova,
with observations across
much of the electromagnetic spectrum and possibly via other messengers.  But the
question remains as to how many were
missed in the historical past, when the naked eye was the only available
detector.

In this paper we study the naked-eye visibility of supernovae in our Galaxy, building in particular
on the work of \citet{adams}.
In \S \ref{sect:SNdist} we develop a general formalism to describe the supernova probability distribution in space and on the sky.  This depends on the supernova spatial distribution, and we adopt axisymmetric models for
this and for the Galactic dust distribution.
In \S \ref{sect:SNvis} we gather the information
needed to assess the naked-eye visibility of Galactic supernovae.  We use recent extragalactic supernova light curve data to assess the sustained luminosity of different
supernova subtypes.  We also model dust extinction effects,
and review data on historical supernovae and present-day Galactic supernova rates.  In \S \ref{sect:results}
We present our model results, first showing sky maps of
the intrinsic supernova probability on the sky.  Then we show the sky distributions for naked-eye supernovae and compare them to the location historic events.  We discuss in \S \ref{sect:discuss} the implications of the surprising comparison with historical data. We conclude in \S \ref{sect:conclude} by summarizing our results and 
suggesting future work.




\section{Galactic Supernova Distributions}

\label{sect:SNdist}

\subsection{Formalism}

Our calculations are grounded in the spatial distribution of supernovae in the Galaxy.
We develop a general formalism, but then construct a model that has three basic assumptions. The most drastic of these is that our Galaxy can be approximated as spatially axisymmetric, which means that we are ignoring the effects of both the central bar and spiral arms of the Milky Way galaxy. We also assume that both stars and dust
can be characterized by exponential distributions in both height above the Galactic plane and radius from the central axis. Finally, we assume that over the timescales of interest the average rate of Galactic supernovae does not evolve;
this is certainly the case for 
the historical period,
where the larger uncertainty
is due to the stochasticity of the
precious few events for which we have
records.
We plan future work in which we relax some of these assumptions, but we also note that the are
not uncommon and were also adopted by \citet{adams}.

The fundamental quantity we use is the rate density $q_i$ for supernova of type $i$:
\begin{equation}
    \frac{d{\cal N}_{i}}{dV \ dt} = q_i(R,z)
\end{equation}
which gives the number of events per unit volume and time.
Integrating the rate density over the entire volume  gives
the total Galactic rate of $\RSN_i=\int q_i \ dV$ of supernovae of type $i$.  We can go on to write a probability distribution function, 
\begin{equation}
    \rho_i(R,z) = \frac{dP_{i}}{dV} = \frac{q_i(R,z)}{\RSN_i}
\end{equation}
which gives the probability per unit volume of finding
a supernova of type $i$ at a point $(R,z)$. This probability density
distribution is properly normalized, so that integrating over the 
Galactic volume gives $\int \rho \ dV  = 1$.
It also follows that we have $q_i = \RSN_i \ \rho_i$.

We will follow a common approach to modelling the Galactic disk components, approximating both radial and height density as exponential drop-offs.  In this ``double exponential'' model,
the disk component $j$ has a normalized spatial probability distribution
\begin{equation}
    \label{eq:density}
    \rho_j(R,z) = 
    \frac{\exp(-R/R_{j}) \ \exp(-|z|/h_j)}
    {4\pi R_{j}^2 h_j}
\end{equation}
where the two free parameters are the scale length $R_j$
and scale height $h_j$.
In this paper we will adopt distributions of this form,
as did \citet{adams}.

 We start with our basic quantity: the supernova probability per unit area of the sky as seen from the Sun's location at a Galactocentric radius and height of $(R_\odot,z_\odot)$. 
We use the observer's Sun-centered Galactic (spherical) coordinates
to calculate the supernova probability per solid angle, i.e., probability density per angular area.  Namely, along a sightline in the $(\ell,b)$ direction, the
probability density out to a distance $r_{\rm max}$ is
\begin{equation}
\label{eq:dP/dOmega}
    \frac{dP_i}{d\Omega}(r_{\rm max},\ell,b)
    = \int_{r=0}^{r_{\rm max}} r^2 \ \rho_i(R,z) \ dr
\end{equation}
which follows from the spherical volume element $dV = r^2 \, dr \, d\Omega$
with solid angle $d\Omega = \cos \ell \ d\ell \, db$
in Galactic coordinates.
Inside the integral, the density depends on Galactocentric cylindrical coordinates given by
\begin{eqnarray}
\label{eq:R}
R(r,\ell,b) & = & \sqrt{R_\odot^2 + r^2 \cos^2 b - 2 R_\odot r \cos l \, \cos b} \\
\label{eq:z}
z(r,\ell,b) & = & r \, \sin b + z_\odot
\end{eqnarray}

We find the maximum distance $r_{\rm max}$ in each direction by imposing a limiting apparent magnitude
$m_{\rm lim}$ (i.e., observed flux) for
the supernova.  
The sightline distance follows from this limiting magnitude, together with the supernova absolute magnitude (i.e., luminosity) and the dust extinction, as explained in detail below in \S \ref{sect:SNvis}.
Thus the probability of finding a SN of type $i$ with a limiting magnitude $m_V^{\rm lim}$ or brighter
is
\begin{equation}
\label{eq:dP/dOmega_mlim}
     \frac{dP_i}{d\Omega}(m_V^{\rm lim},\ell,b) 
     = \frac{dP_i}{d\Omega}
     \left[r(m_V^{\rm lim}),\ell,b \right]
\end{equation}
where $r(m_V^{\rm lim})$ is solution to $m_V(r) = m_V^{\rm lim}$.
And finally, the total probability of finding type $i$ events is just
\begin{equation}
\label{eq:Ptot}
    P_i(m_V^{\rm lim}) 
    = \int \frac{dP_i}{d\Omega} \ d\Omega
    = \int \frac{dP_i}{d\Omega} \ 
    \ \cos b \ db \ d\ell \ \ .
\end{equation}

Equations (\ref{eq:dP/dOmega}-\ref{eq:Ptot}),
are quite general--they hold for any axisymmetric distribution, and are straightforward to extend to non-axisymmetric cases.  
To evaluate them, we now 
build a model using the approximations noted above, turning first to the spatial distributions.

\subsection{Models for Milky Way Stars, Dust, and Gas} 

Our fiducial model for the spatial distribution of supernova populations and dust in our Galaxy follows \citep{adams}, which uses
the results of the TRILEGAL model \citep{trilegal}
that uses stellar data to fit both a thin and thick disk,
both in the double exponential form of eq.~(\ref{eq:density}) above.
The two components have scale parameters
\begin{eqnarray}
\label{eq:thindisk}
(R_{\rm thin},h_{\rm thin}) & = & (2.9 \ \rm kpc, 95 \ \rm pc) \\
\label{eq:thickdisk}
(R_{\rm thick},h_{\rm thick}) & = & (2.4 \ \rm kpc, 800 \ \rm pc)
\end{eqnarray}
In addition, TRILEGAL adopts a dust distribution 
\begin{equation}
\label{eq:dustdist}
    (R_{\rm dust},h_{\rm dust}) = (2.9 \ \rm kpc, 110 \ \rm pc)
\end{equation}
that is nearly identical to the thin disk, reflecting a common
origin in cold, star-forming regions.

We follow the \citet{adams} associations of supernovae with these
disk components, as summarized in Table \ref{tab:sn_types}.  We model core-collapse SNe with the
thin disk distribution above:  $\rho_{\rm CC} = \rho_{\rm thin}$.  We model Type Ia supernovae as 
having two components:  $\rho_{\rm Ia} = (\rho_{\rm thick}+\rho_{\rm thin})/2$:  
half arise in the thick disk, and half in the
thin disk, corresponding to long and short delay times.

Two assumptions are worth highlighting.
The TRILEGAL model, and thus ours, assumes an axisymmetric galaxy.
This omits Milky Way features such as a bar and spiral arms.
These may be important, as we will see below, and
we plan to study their effects in future work.
The last assumption we use is the time frame we are considering for supernovae is small enough to where the rate of supernovae occurring remains constant; this is certainly appropriate for historical supernovae.

\begin{table} 
    \centering
        \caption{Properties of different types of supernovae}
    \begin{tabular}{c||cc}
        \hline \hline
        Property & \multicolumn{2}{c}{Supernova Type} \\
        \hline
        Mechanism & Core-Collapse & Thermonuclear \\
        Observed Type & II-P, II-L, Ib/c & Ia \\
        Galactic Tracer & Thin Disk & Thick and Thin Disk \\
        MW Rate${}^{a}$ [events/century] & $3.2^{+7.3}_{-2.6}$ & $1.4^{+1.4}_{-0.8}$  \\
        \hline \hline
    \end{tabular}
    \label{tab:sn_types}
    \mbox{${}^{a}$ From \citet{adams}}
\end{table}

Other ways to model the supernova distributions are possible. 
A full inventory of such approaches is beyond the scope of this paper, but here we highlight two.
\citet{Green2015} compiled a Galactic supernova remnant catalog (also discussed in \S \ref{sect:SNcats} below),
and used their longitude distribution to constrain the underlying supernova profile in 
Galactocentric radius.
For an exponential form, the best-fit scale radius
is $R_{\rm SNR} = 3.1 \ \rm kpc$, based on a fit that omits the $|\ell| < 10^\circ$ region where
radio fore/backgrounds likely bias against remnant identification.  The \citet{Green2015} catalog includes both CCSN and SNIa, and so this represents
a sort of average of the two; the TRILEGAL model we use has a very similar value for the
thin disk
(eq.~\ref{eq:thindisk}) but a slightly smaller value for the thick disk (eq.~\ref{eq:thickdisk}).
We thus expect our CCSN distribution to be similar to that of the SNR catalog 
regarding the longitude distribution, while our SNIa distribution reflects somewhat a more radially compact thick
disk.

In order to compare our results with a dramatically different case, we consider
the Galactic distribution of \al26 from \citet{Wang2020}.
This study presents an updated analysis of the sky distribution of the 1.809 MeV nuclear
gamma-ray line from the radioactive decay of \al26.  \al26 production is likely dominated by
massive stars, with SNIa production expected to be small.
Thus the gamma-ray line emission should trace Galactic CCSN activity as averaged over the
decay mean life $\tau(\al26) = 1.0 \ \rm Myr$. Fitting the \al26 sky to a double exponential
as in eq.~(\ref{eq:density}), \citet{Wang2020} find scale parameters 
\begin{equation}
\label{eq:26Al}
(R_{{}^{26}{\rm Al}},h_{{}^{26}{\rm Al}}) = (7.0^{+1.5}_{-1.0} \ {\rm kpc},800^{+300}_{-200} \ {\rm pc}) 
\end{equation}
Remarkably, the scale radius is more than twice that of our thin disk value in eq.~(\ref{eq:thindisk}).
Moreover, the scale height is eight time larger that than our in our fiducial thin disk ,
and indeed is the same as our adopted {\em thick} disk value of eq.~(\ref{eq:thickdisk}).

We will adopt the central values in eq.~(\ref{eq:26Al}) to illustrate the effects of these
large differences on our results.\footnote{\citet{Wang2020} also fit to the sky distribution of the \iso{Fe}{60} gamma-ray decay lines, also finding a large scale radius and height.  But
the authors caution that these results are less reliable for this dimmer signal. } 
In comparing this distribution to our fiducial model, it is well to bear in mind two caveats.
First, if \al26 has significant nucleosynthesis sources other than massive stars,
these will affect the resulting sky and spatial distributions. For example,
asymptotic giant branch stars are likely \al26 producers \citep[e.g.,][]{Doherty2014}, and this longer-lived
population could have a thicker height distribution. 
Second, \citet{Fujimoto2020a} have suggested that the \al26 sky map is substantially perturbed 
by foreground emission due to the Sun's embedding in a local source of \al26 nucleosynthesis.
This intriguing result is consistent with other evidence of recent near-Earth supernovae,
notably the detection of \iso{Fe}{60} as summarized in \citet{Fields2019}.
On the other hand, \citet{Krause2020} simulate the Galactic disk and find some \al26 is vertically dispersed as high as $\sim 50 \rm kpc$ in the halo.
With these caveats in mind, the \al26 distribution may represent a sort of upper limit
to the possible lengthscales of the Galactic supernova distribution.

\subsection{Integration Methods}

We used two different methods to model the probability and visibility of supernovae across the sky,
i.e., to solve eqs.~(\ref{eq:dP/dOmega_mlim})
and (\ref{eq:Ptot}).  The first method, described in further detail in Appendix \ref{app:newton}, uses Newton's method to calculate how far along the sightline an observer can see before extinction causes the light to be too dim to see. The probability distribution of supernovae is then integrated along each sightline to produce ${dP}/{d\Omega}$.
The 2D sky plots below are generated with this approach.

The second method, described in Appendix \ref{app:MC}, is a Monte-Carlo integration. We generate a large number ($>10^4$) of points representing supernovae, distributed according to the 
appropriate density function. The extinction and distance modulus of each point is then calculated and compared with the difference between the absolute magnitude of the SN type and the limiting visible magnitude. Their apparent magnitude is then compared across a range of limiting apparent magnitudes to compute $P_i(m_V)$
and thereby generate Figure \ref{fig:viz_vs_mag}.

\section{Supernova Observability:  Luminosities, Historical Events, and Rates}
\label{sect:SNvis}

\begin{figure}
    \centering    \includegraphics[width=\columnwidth]{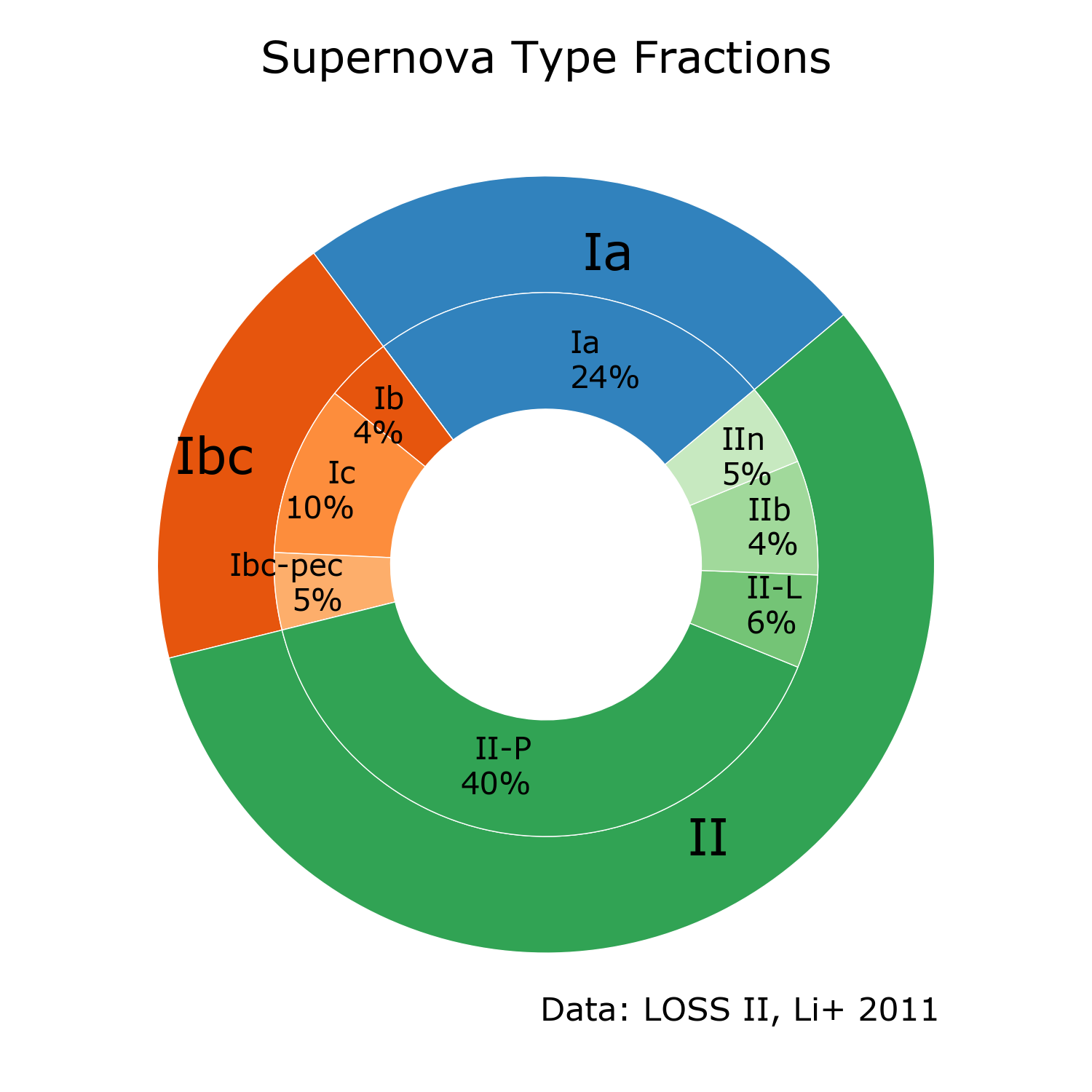}
    \caption{Intrinsic rates of supernovae: relative fractions
    of different types and sub-types as determined in
    a volume-limited sample in the nearby universe.  
    From the \citet{Li2011} analysis of the LOSS survey data
    for 175 supernovae.}
    \label{fig:SNTypes}
\end{figure}

The observability of a Galactic supernova depends
on its intrinsic luminosity, its distance $r$ from Earth,
and on dust extinction along the line of sight.  
Specifically, 
a 
supernova with $V$-band absolute magnitude $M_V$ and located at a distance $r$ from Earth, will be observed with an apparent magnitude
\begin{equation}
\label{eq:appmag}
    m_V(r) = M_V + \mu(r) + A_V(r) \ \ .
\end{equation}
Here
$\mu(r) = 5 \log_{10}(r/10 \, \rm pc)$ is the distance modulus, and $A_V(r)$ is the extinction along the sightline.
In this section we consider in turn 
$M_V$ and its relation to the supernova light curve,
and $A_V$ and its relation to dust along the sightline.

\subsection{Naked-Eye Supernova Discovery}

For historical supernovae, we are interested in the observability to the naked eye.  We  use
 the $V$ passband as a proxy for the visible range.
The naked eye has a sensitivity of at best $V_{\rm lim}^{\rm eye} \simeq 6 \ {\rm mag}$ on a dark moonless night without light pollution \citep{Clark1990,Cinzano2001}.\footnote{In modern times, this corresponds to a class of 3 to 4 on the Bortle scale \citep{Bortle2001}.}  
We expect a supernova must appear substantially brighter that this in order
to be discovered confidently and to be deemed worthy of
historical note.  \citet{Clark1977} quote a minimum brightness for detection of $m_V = 3$ in their Fig.~1.5.  
To our knowledge most magnitude estimates for historical supernova observations are brighter than this, though very difficult to 
estimate accurately \citep{Stephenson2002}.\footnote{
For a modern comparison, SN 1987A
was discovered visually at $m_V \simeq 5 \ \rm mag$,
by several groups of professional astronomers
who had been monitoring the Large Magellanic Cloud 
from a variety of observatories \citep{Kunkel1987}.  
}  
In the context of the non-observation of Cas A, \citet{Shklovsky1968}
suggest that ``this outburst would hardly have
remained unnoticed if its visual magnitude had been 
brighter than'' $m_V = 4$.   \citet{vandenBergh1970}
argue for a higher threshold of $m_V = 2$, based
on a comparison between 
the modern observed bright nova rate versus
the sparser Chinese records of guest stars.

For a fiducial threshold apparent brightness for historical events,
we adopt\footnote{Here we are grateful for private discussions with Ralph Neuh\"auser.}
the most conservative (i.e., brightest) of these values,
\begin{equation}
    m_{V,\rm lim}^{\rm SN} \simeq 2 \ {\rm mag} \ .
\end{equation}
The detection threshold is clearly not sharply defined,
so we will
show how our results change for different values of 
$m_{V,\rm lim}^{\rm SN}$.




\subsection{Supernova Luminosities}

\label{sect:luminosity}

The
peak luminosity corresponds to the 
peak absolute magnitude, $M_V^{\rm peak}$;
while this sets an important scale,
strictly speaking the explosion is only instantaneously this luminous.  
For a Galactic supernova to be identified
and to enter the historical record requires repeated observations.  This demands that the event
is visible over an extended period of time
so that observers would have confidence that a ``new'' or ``guest'' star is real.
Difficulties with weather and observing season can
lengthen the time needed for such observations.  To attempt to capture this, we define $M_V(\Delta t)$ to be
the absolute magnitude at which the $V$-band-light curve has width $\Delta t$.  In other words, over the timescale $\Delta t$ the supernova is at least this luminous: $M_V \leq M_V(\Delta t)$.  We thus call this a {\em sustained luminosity} for interval $\Delta t$.  For a singly-peaked supernova light curve, the shorter $\Delta t$, the closer the sustained magnitude is to the peak absolute magnitude.
The {\em minimum} duration $\Delta t$ required for historical supernova
detection is of course uncertain,
and so we determine sustained luminosites for
a range of durations.

\begin{center}
\begin{table*}
        \caption{Supernova $V$-band absolute magnitudes:  peak and sustained luminosities}
    \begin{tabular}{cc|cccccl}
    \hline \hline
        SN & & Peak & \multicolumn{4}{c}{Sustained $M_V(\Delta t)$} & \\ 
        Type & Template & $M_V^{\rm peak}$ & $\Delta t = 30 \ \rm days$ & 60 days & 90 days & 180 days & References \\
        \hline 
        Ia & 2011fe & -17.7 & -16.9 & -15.3 & -14.7 & -12.8 &  
        \citet{Munari2013,Brown2014,Stahl2019}
        \\  
        II-P & 2012aw & -17.0 & -16.8 & -16.7 & -16.5 & -13.9$^*$ &
        \citet{Munari2013,Brown2014,DallOra2014}  
        \\  
        II-L & 2014G & -17.1 & -16.3 &-15.6 & -14.2 & -11.8$^*$ &
        \citet{deJaeger2019}
        \\  
        IIb & 2011dh & -16.8 & -16.5 & -16.0 & -15.3 & -12.3$^*$ & \citet{Sahu2013,Bartunov2007}\\
        IIn & 2011ht & -16.8 & -16.5 & -16.1 & -15.6 & -10.2$^*$ & \citet{Mauerhan2013,Bartunov2007,Brown2014}\\ 
        Ib & 2016coi & -17.5 & -16.0 & -15.3 & -14.9 & -13.4 & \citet{Prentice2018,Brown2014}\\
        Ic & 2007gr & -16.7 & -15.3 & -14.5 & -14.0 & -12.3 & \citet{Bartunov2007,Bianco2014}\\
        \hline \hline
    \end{tabular} \\
    \mbox{$^*$Estimated \hfill \hphantom{m}}
    \label{tab:SNabsmag}
\end{table*}
\end{center}

Table \ref{tab:SNabsmag} shows sustained luminosity values
for major supernova varieties.
These results drew from the data compilation
and visualization in the the Open
Supernova Catalog\footnote{\href{https://sne.space/}{https://sne.space/}} \citep{Guillochon2017}.
For each supernova type or subtype, we chose a
``template'' or exemplar event 
that: (a) showed consensus in the type or subtype classification,
(b) had among the largest apparent 
brightness, (c) was well observed in the $V$ band for
at least 180 days, and (d)
was seen within the last 20 years.
%
%
For each event, we found
$M_V^{\rm peak}$ and $M_V(\Delta t= \rm 30, 60,$ \& 90 days).
As noted in \citet{Li2011}, intrinsic variations
within the different subtypes are typically $\sigma(M) \sim 1$ mag.
Also, we have not attempted to correct for extinction in the
host galaxy; these events were generally in the outskirts
of face-on galaxies and where estimates exist,
they are less than 1 mag.  

In Table \ref{tab:SNabsmag} we see that the sustained luminosities
can be much smaller than the peak values.  For example, $M_V(90 \, \rm days)$ can be 2 magnitudes dimmer
than the peak values.
Turning to specific types,
SNIa have the highest peak luminosity,
but at $\Delta t  \ge 60$ days,
the core-collapse events have higher
sustained luminosity.  This
reflects difference in the light curve shapes
over these timescales.

For the core-collapse events, we see the the
most common subtype, II-P (``plateau''), have the highest
sustained luminosity $\ge 60$ days,
due to the extended plateau of brightness
that gives them their names.
We also see that the spread among the
core-collapse subtypes grows
from 0.8 mag at peak to
3.7 mag at 180 days.
In the results below
we adopt the 
Table \ref{tab:SNabsmag} 
value for SNIa, while for
CCSN, we take a weighted average of each according to the relative abundances shown in Figure \ref{fig:SNTypes}.

\citet{Stephenson2002} show that 
the 5 confirmed historic events
were visible for timescales ranging
from 6 months (SN 1181) to 3 years for SN 1006.
Thus a choice of 90 days visibility would
be shorter than all known historical supernovae,
though presumably enough time to allow for detection.
We will adopt $M_V({\rm 90 \ \rm days})$ 
for which we have the best data,
but will compare with results
assuming $M_V({\rm 180 \ \rm days)}$.

We assume all core-collapse subtypes follow the same 
thin-disk spatial distribution.  This implies that the different subtypes occur 
with the same relative frequencies throughout the Galaxy.
If this were not the case, it could affect our results
if the intrinsically brighter events were concentrated
differently than the rest.
In fact, \citet{Kelly2008} studied location of 
504 low-redshift supernovae in their host galaxies.
The found that Type Ia, Ib, and II events trace the galaxy light,
while Type Ic events are preferentially located in the brightest regions of their hosts.  Later work shows that Type IIb and broad-line Type Ic and events favor the bluest regions with indications of lower metallicity
\citep{Kelly2012}.
Our work does not capture these complex effects.

\subsection{Dust Extinction}

We calcuate the extinction in our Galaxy in a heliocentric coordinate system. The optical depth in the $V$ passband 
is the integral over a line of sight:
\begin{equation}\label{7}
    \tau_V = \kappa_V \int_{\rm los}\rho \, ds = \kappa_V \Sigma = \frac{2 \, \ln 10}{5} A_V
\end{equation}
with $\Sigma=\int_{\rm los} \rho \, ds$ the mass column along the sightline, and $\kappa_V$ the opacity.
This integral in general is not simple since the density along each individual line of sight varies throughout the sky, and thus we must evaluate it
numerically.  

We normalize to the extinction
sight line of Galactic center, which
we take to be $A_{V,\rm GC} = 30$
following \citet{adams}.
This gives
\begin{equation}\label{eq:AV}
    A_V(r_{\rm max},\ell,b) 
= \frac{\int_{0}^{r_{\rm max}}\rho_{\rm dust}(R,z) \ dr}{\Sigma_{\rm GC}}
A_{V,\rm GC}
\end{equation}
where $r_{\rm max}$ is the same as in eq.~(\ref{eq:dP/dOmega}).
The denominator in eq.~(\ref{eq:AV}) is the mass column $\Sigma_{\rm GC}$ to the Galactic Center;
for our case of axisymmetry with an exponential radial drop-off, this is
$\Sigma_{\rm GC} = \rho_{\rm dust,0} R_i(1-e^{-R_{\odot}/R_i})$.
Equations (\ref{eq:R}) and (\ref{eq:z}) give the
conversions from heliocentric sightline coordinates
to the Galactocentric coordinates that appear in 
the dust distribution.

Our simple double-exponential model for the Galactic dust distribution is only a rough approximation.  \citet{adams} showed the effect of using more sophisticated dust distributions based on observed extinction maps. They found that the qualitative results for Galactic supernova visibility were similar,
and that generally results for this simple model fell between those based on those of \citet{Schlegel1998} and a modified version with lower $E(B-V)$
reddening similar to that in updated later work \citet{Schlafly2010,Schlafly2011}.  Subsequently, 
\citet{GreenG2014} and \citet{GreenG2019} presented
full 3D dust maps based on PAN-STARRS and 2MASS data \footnote{\href{http://argonaut.skymaps.info/}{http://argonaut.skymaps.info/}}.
These very useful tools have some limitations for our purposes,
with accuracy limited to distances $< 10 \ \rm kpc$ for the low Galactic latitudes of most interest to us.  Nevertheless, it would be of great interest to incorporate these maps in future studies.

\subsection{Data:  Supernova Remnants and Historical Supernovae}
\label{sect:SNcats}

We will calculate both the intrinsic (all brightness)
sky distributions of supernovae,
as well as the expected sky map of naked-eye events.
We will compare these predictions with observed
events.  To represent the full
complement of known Galactic supernovae,
we use the
\citet{Green2019}
catalog of 294 Milky-Way supernova remnants.
Most of these objects do not have secure
distances or type designations.
The catalog is based largely on
radio detections,
so should not suffer from extinction
effects, due to the Milky Way
transparency in the radio. However,
as \citet{Green2015}
elucidates in detail,
the available radio surveys
are highly nonuniform for several reasons.
They have have a lower sensitivity to SN remnants
in inner Galaxy,
due to the high radio background in these regions.
Moreover, northern radio surveys
have been more numerous and deeper
than their southern counterparts.
Also, additional biases could arise if
supernova remnant lifetime and brightness
varies with location in the Galaxy,
as is likely.



Turning to historic supernovae, we draw upon 
the extensive studies in the \citet{Clark1977}
and \citet{Stephenson2002}
monographs, and subsequent update
\citet{Green2017}.
There are 5 confident historic supernovae on record, occurring in the years 1006, 1054, 1181, 1572, and 1604. SN 1006, the oldest supernova on record, is also the southernmost and was likely the brightest as well ($\mVpeak \approx -7$). At its peak, it was easily seen during the day and was tracked by astronomers for 3 years.
SN 1054, whose remnant is commonly known as the Crab Nebula, is the second oldest and most studied, being
extensively cataloged by Chinese and Japanese astronomers.
Other than SN1006, it is the only SN that was recorded as being seen during the day, and was visible for 21 months,
with an estimate peak brightness $\mVpeak \approx -3.5$. Remarkably, 
SN 1054 is located at almost the anticenter of the Milky Way, which will be important in our results below. 
SN1181 was seen by Chinese, Japanese, and Arab
astronomers for 6 months, its
peak brightness is estimated at $\mVpeak \simeq 0$ or perhaps
slightly higher \citet{Stephenson2002}.
SN 1572, also known as Tycho's supernova, was seen for 18 months and attained a peak brightness of about $\mVpeak \approx -4$
\citep{Stephenson2002}.
The last supernova seen in the Milky Way was SN1604, which was seen by Chinese, Korean, and European astronomers, including
most famously Johannes Kepler. \citet{Clark1977} 
infer a rough light curve that spanned 12 months, and find a peak magnitude around $\mVpeak \approx -3$.

For these confirmed supernovae, the basic types are 
confidently established for four the five.
Both SN 1054 and 1181 were core-collapse events, as confirmed by
the observation of a pulsar in each remnant.
The Crab is thought to be IInP \citep{Smith2013}, which would be a signature of interactions with a circumstellar medium.
SN 1181 seems to be underluminous and with circumstellar material,
but even so the type is unclear:  \citep{Kothes2013} proposes
Ib, Ic, II-L, or and underluminous II-P.
SNe 1006, 1572, and 1604 are identified as Type Ia events.
For SN 1006 this is primarily 
because of its location outside of the
Galactic plane, lack of nearby OB association, and
high luminosity and lack of compact object in the remnant
\citep[][and refs.~therein]{Katsuda2017}.
In the case of Tycho this has been confirmed by light echo spectroscopy \citep{Krause2008}.
For Kepler this
based on their remnant composition and lack of a compact object
in the remnant \citep[][and refs.~therein]{Vink2017}.

It is worth noting that all five confirmed historical supernovae
are visible from Earth's northern hemisphere, as seen
below in Fig.~\ref{fig:SN-summary}.  \footnote{All but the Crab are located in the northern Galactic hemisphere, as also seen
in Fig.~\ref{fig:SN-summary}.  Presumably this arises because of the partial overlap between the Galactic and equatorial north.}.
We suspect this is a selection effect, reflecting that fact that
most searches to date focused on
 civilizations in the northern hemisphere, searching Chinese, Japanese, Korean, Indian, Arabic, and European records.  We expect that interesting results await searches in records from the global South.  
 
 Indeed, the Galactic Center is best observed from the south.  This may play a role in
the curious paucity of known historical events there, which
 we will see contrasts with our model predictions that this should be the region of maximum probability.
 
 Several additional events 
 or remnants have been proposed 
 as historical SNe, but 
are not (yet) supported by the 
same weight of
 evidence of the five `confirmed' events.
We will highlight the
three of these that
\citet{Green2017a}
 consider likely,
 and that are associated
 with specific remnants:
 Cas A, SN 185 and SN 386.
The Cas A remnant
has an associated pulsar
and thus it
was a CCSN.  As noted above,
although there is some debate,
it seems unlikely Flamsteed saw
this event and thus we do not count
it among confident/confirmed historical
SNe.

Chinese records 
describe a guest star in AD 185,
whose
position is most likely
associated with the
G315.4-2.3 (RCW 86)
remnant
whose age is consistent
\citep{Vink2006}.
The remnant location is 
consistent with an OB association,
and thus suggests an origin in
a core-collapse explosion
\citep{Vink1997}. 
Similarly, one Chinese source
records a guest star in 
AD 386, which has been associated 
with the G11.2-0.3 remnant
that contains a pulsar and
thus was a CCSN.

\subsection{Supernova Rates and Frequencies}

A number of methods have been used to estimate
the present rates of supernovae in the Milky Way.  
\citet{adams} use historic supernovae
and their calculations of naked-eye observability
to infer global rates for the two main supernova types.
These appear in Table \ref{tab:sn_types},
where we see that
core-collapse events occur at about double the
rate of Type Ia explosions.
Furthermore their total Galactic rate
of both types is ${\cal R}_{\rm CC} + {\cal R}_{\rm Ia} = 4.6^{+7.4}_{-2.7} \ \rm events/century$, so that 
the $1\sigma$ range spans 
a mean interval of $\sim 8$ to $50$
years per event.  Thus we can
reasonably hope for a event somewhere in 
the Galaxy within
a human lifespan.


Recently, \citet{Rozwadowska2021}
survey this and other methods of estimating Galactic CCSN
rates, and in a combined analysis they 
find ${\cal R}_{\rm CC} = 1.63 \pm 0.46 \ \rm events/century$.
This is consistent within errors with the \citet{adams} result,
but reports a central value that is almost a factor of 2 smaller,
and a significantly smaller error. 

Core-collapse events display a range of observational behaviors
that have been grouped into subtypes.  The intrinsic frequencies 
for these subtypes are summarized in 
in Fig~\ref{fig:SNTypes},
which presents the data from the 
\citet{Li2011} volume-limited analysis of the
LOSS-II survey.  We see that Type II-P 
supernovae
are the most common core-collapse events.

\section{Results:  Milky Way Supernovae in Axisymmetric Galactic Models}

\label{sect:results}

We now present our predictions for the intrinsic and observed supernova distributions for our axisymmetric model of the Milky Way supernova distribution.

\begin{figure*}
    \centering
    \includegraphics[width = \textwidth]{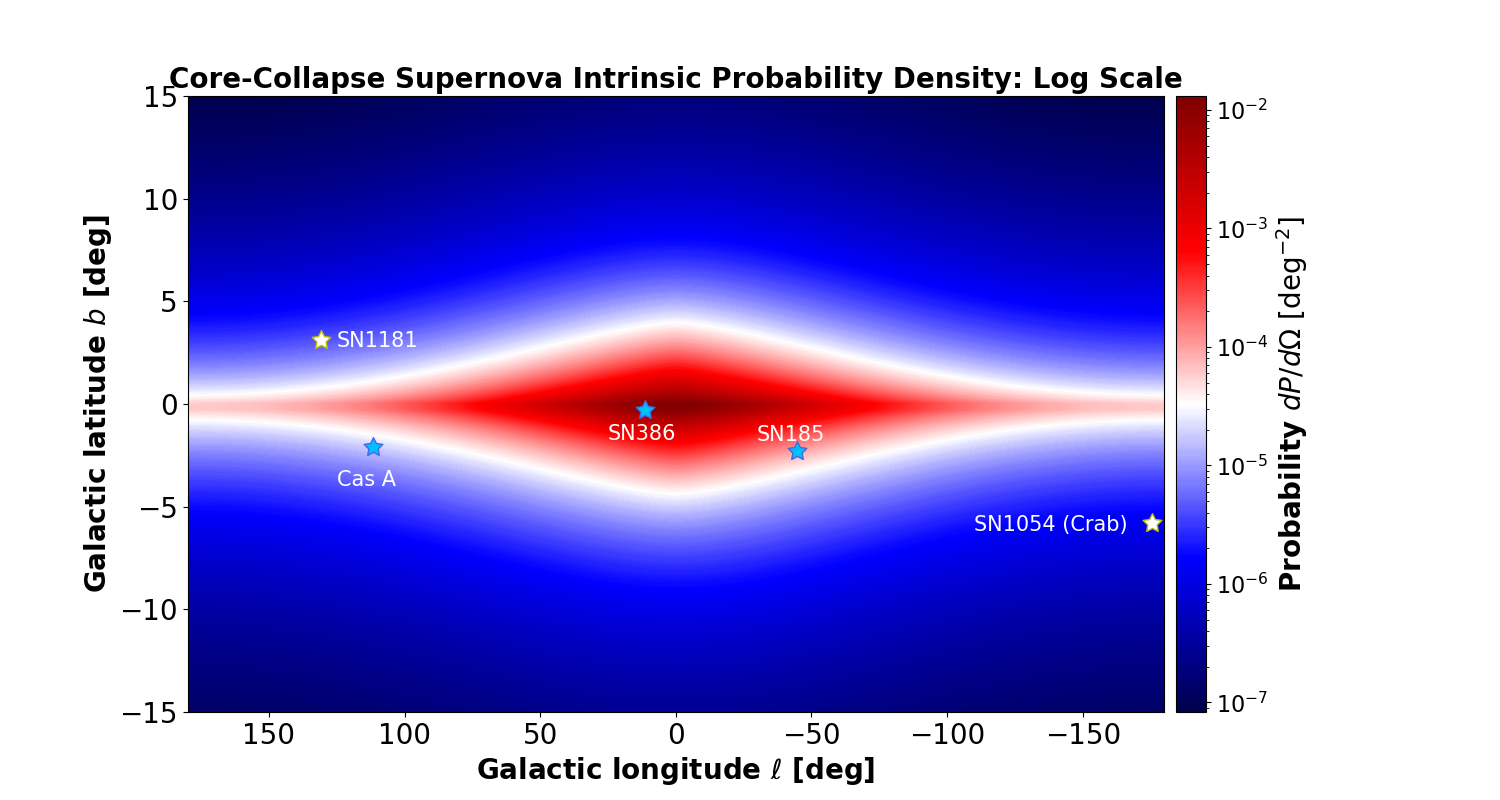}
    \caption{Sky plot of the probability 
    density for all Milky Way 
    core-collapse supernovae,
    in units of probability per square degree;
    {\em note the logarithmic scale}.
    Shown in Galactic coordinates is a
    zoom into the region around the Galactic
    plane spanning all longitudes but
    only latitudes $|b| \le 15^\circ$. 
    Note that the vertical dimension is
    stretched compared to the horizontal
    in order to highlight the interesting regions.
    The intrinsic probability density is shown, that is,
    the map plots the likelihood
    for sky positions for all CCSN 
    to occur,
    regardless of distance, apparent magnitude 
    or extinction effects.}
    \label{fig:cc_wo_ex}
\end{figure*}

\begin{figure*}
    \centering
    \includegraphics[width = \textwidth]{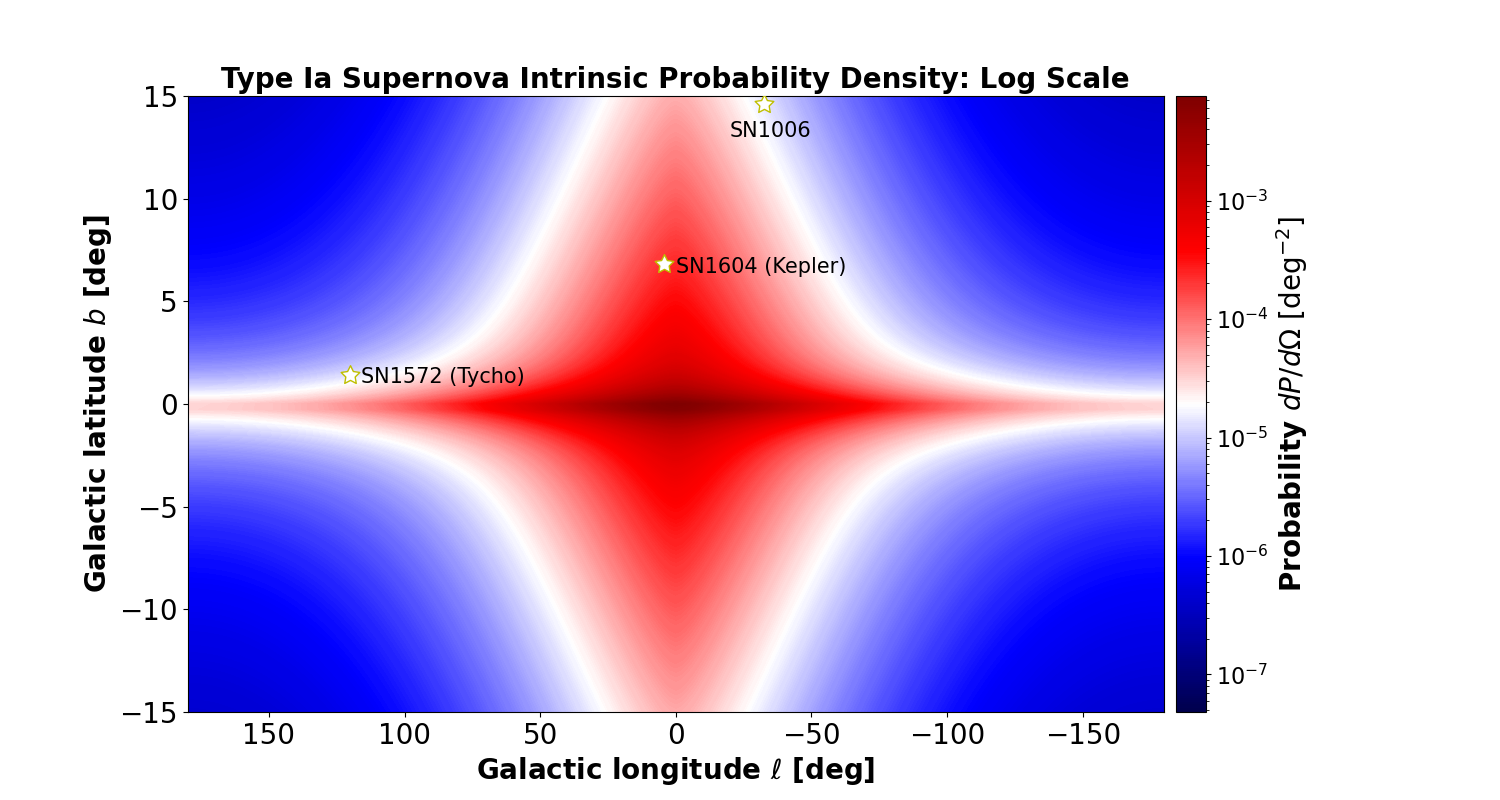}
    \caption{Similar sky plot to \ref{fig:cc_wo_ex}, but for the intrinsic distribution of Type Ia supernovae. {\em Note the logarithmic scale}.}
    \label{fig:Ia_wo_ex}
\end{figure*}

\subsection{The Intrinsic Galactic Supernova Sky Distribution}

We begin with the intrinsic sky distributions 
for both core-collapse and Type Ia supernovae. That is,
the probability per unit area on the sky that there will
be a supernova of this type, regardless of its observability.
To do this
we use eq.~(\ref{eq:dP/dOmega}) to calculate $dP/d\Omega(r_{\rm max},\ell,b)$ for each supernova type,
setting the line of sight distance $r_{\rm max}$ to reach the edge of the
Galaxy, which we take to be at $R_{\rm max}=6 R_\odot$; our results
are insensitive to this cutoff choice.

\begin{figure*}
    \centering
    \includegraphics[width=\textwidth]{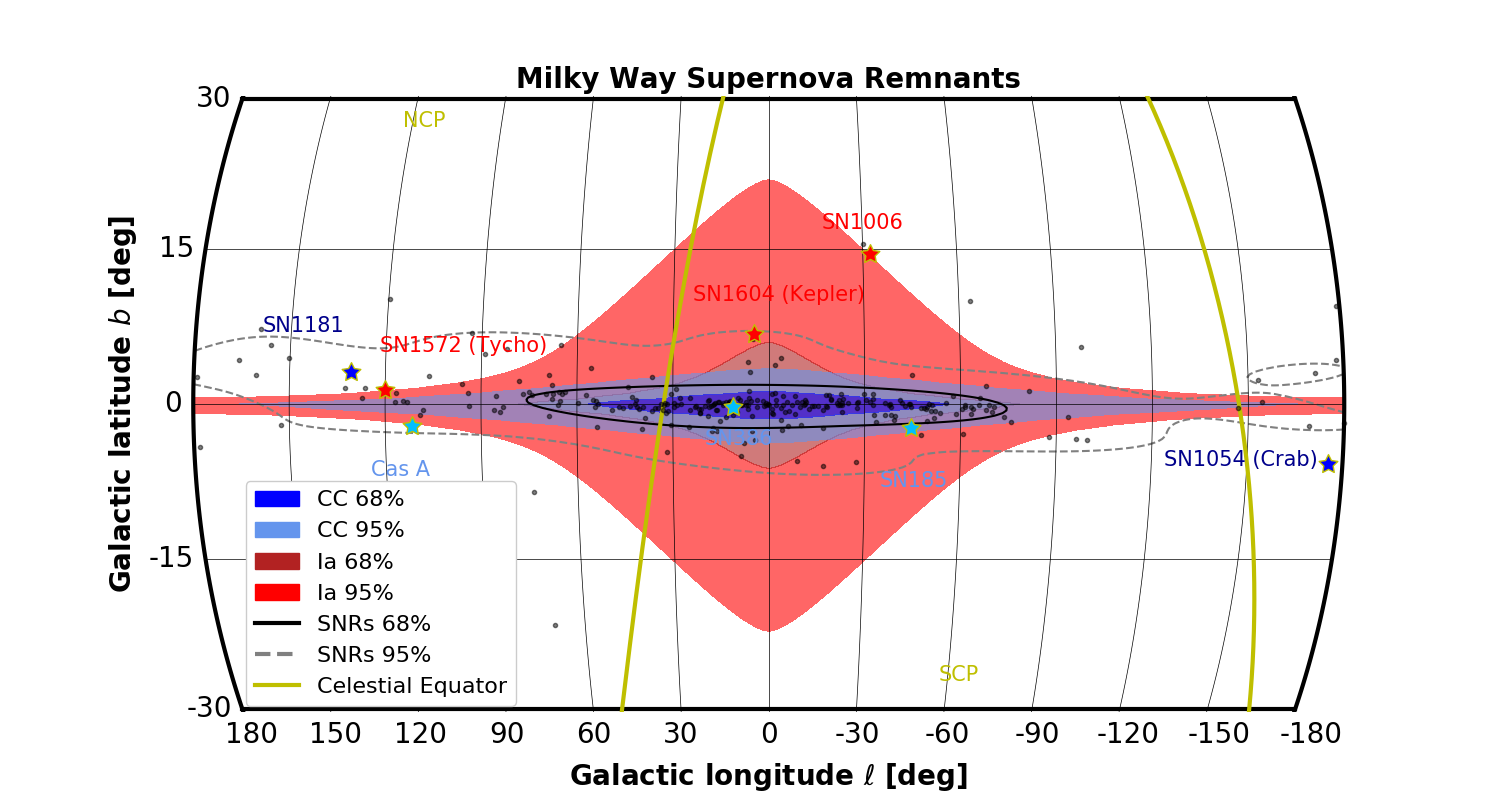}
    \caption{Zoom-out map of supernova probability showing 'zoom-out' views of the distributions
    of CCSN as in Fig.~\ref{fig:cc_w_ex} and Type Ia events as in Fig.~\ref{fig:Ia_w_ex}. 
    In Galactic coordinates
    and Mollweide projection.
    Contours are drawn for $dP/d\Omega|_{\rm contour} = (0.3,1,3) {\rm sr^{-1}}$, for CCSN (blue) as in Fig.~\ref{fig:cc_wo_ex},and SNIa (red) as in Fig~\ref{fig:Ia_wo_ex}.
    Points show the 294 SN remnants
    from the \citet{Green2019}
    catalog; the dark contour lines reflect a kernel density
    estimate and enclose 68\% and 95\% of the events.
    Historical supernovae are labeled: confirmed SNIa events are shown in red;
    confirmed CCSNe are shown in dark blue, while possible CCSNe
    are shown in light blue.
    We see that confirmed historical events
    appear to be outliers in the remnant
    distribution.
    Celestial equator and poles are shown; we see that the asymmetry
    in positive an negative latitudes
    suggesting incompleteness in the third Galactic quadrant.}
    \label{fig:SNR-Sky}
\end{figure*}

Our results appear in Figs.~\ref{fig:cc_wo_ex} and \ref{fig:Ia_wo_ex}, which show the supernova probability 
distribution in the regions of the sky where it is non-negligible, namely near the Galactic plane.  To increase legibility, these and
all sky plots we show are zoomed and {\em stretched} in latitude, i.e., the vertical angular scale is not equal to the horizontal scale.\footnote{An earlier version of these plots appears in \citet{whitepaper}.}
For both supernova types, we see that the region of highest probability is in the central quadrants of the Galaxy
and close to the Galactic midplane.

Core collapse events are confined to within a few degrees of the plane, due to their origin in the thin
disk. The characteristic angular half-width $h_{\rm thin}/R_\odot \sim 0.6^\circ$ compares well with the
thickness of the peak region of Fig.~\ref{fig:cc_wo_ex}.
On the other hand, SNIa in our model have a component in the thick disk that has $h_{\rm thick}/R_\odot \sim 5^\circ$,
which explains the significantly higher vertical extension in Fig.~\ref{fig:Ia_wo_ex}.

For both core-collapse and Type Ia events, there is a subtle shift:  the probability density peak lies not at the
Galactic center itself, but is slightly shifted to negative
latitudes.
This offset to the Galactic south arises due to the
height $z_\odot = 20 \ \rm pc$ of the Sun in the Galactic plane.
so that the sightline to the center is inclined by an angle $\simeq z_\odot/R_\odot \simeq 0.13^\circ$ with respect to the geometric Galactic plane. 
Our vantage point slightly above the midplane means that 
the source density is slightly higher for sightlines 
at negative Galactic latitude than at the same corresponding
positive latitude.  We have run models varying the solar height to show that this is the source of the asymmetry, and to verify that the distribution is symmetric and peaks that the Galactic center when we set $z_\odot =0$.

Our model thus predicts that the
next Galactic SN
is most likely to occur
at low Galactic latitudes 
near the Galactic center,
as one would guess.
Specifically, the CCSN distribution
shows the highest probability
for $|b| \la 1^\circ$
and $|l| \la 45^\circ$.
The heart of the Type Ia distribution
is slightly broader in both dimensions,
with  $|b| \la 2^\circ$
and $|l| \la 60^\circ$.

\begin{figure}
    \centering
    \includegraphics[width = 0.45\textwidth]{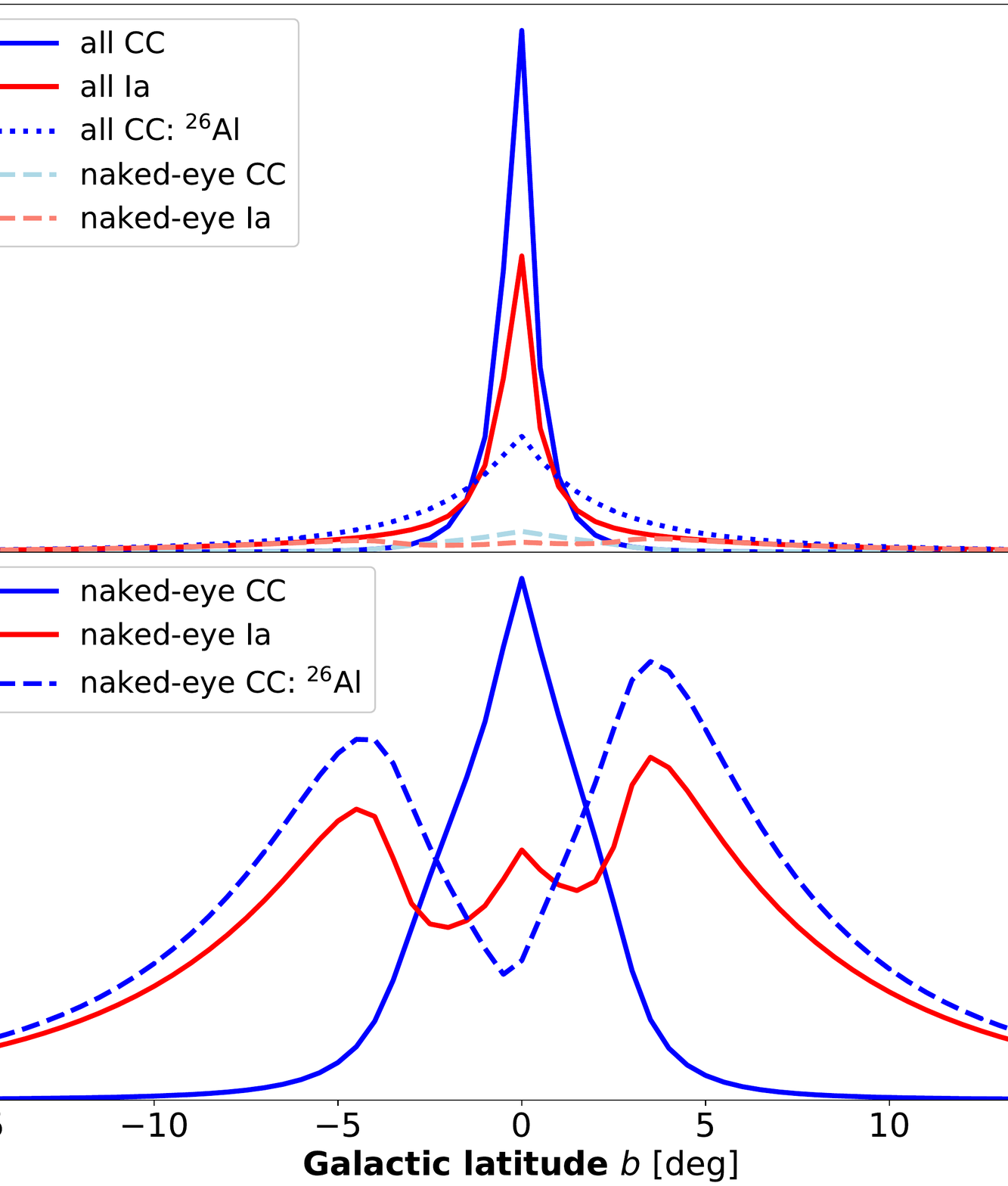}
    \caption{Distribution of Milky Way supernovae in Galactic latitude. {\em Top:} intrinsic and naked-eye distributions
    on the same scale.  {\em Bottom:} zoom into naked-eye distribution.}
    \label{fig:lat-dis}
\end{figure}

\begin{figure}
    \centering
    \includegraphics[width=0.45\textwidth]{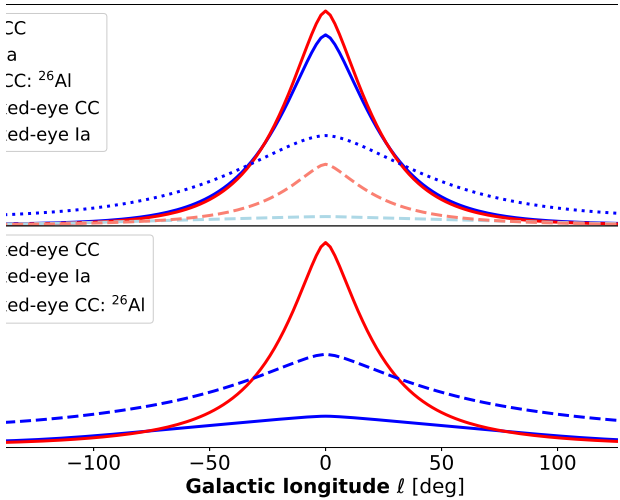}
    \caption{As in Fig.~\ref{fig:lat-dis}, but for supernova longitudes.}
    \label{fig:long-dis}
\end{figure}

Figure \ref{fig:SNR-Sky} 
plots both CCSN and SNIa predictions 
in an zoom-out view.
We also plot the known Milky Way
supernova remnants from
the \citet{Green2019}
catalog.
Because these mostly do not have
well-established types, we do not
attempt to distinguish CCSN vs SNIa remnants
except for historical events.
We see that the remnant distribution
broadly follows the predictions,
with most supernovae found in the central
quadrants and at low latitudes.
We have used a kernel density estimate to derive
a probability density of the remnants;
the solid black (dashed gray) contour in Fig.~\ref{fig:SNR-Sky}
encloses 68\% (95\%) of the observed remnants.

\begin{figure}
    \centering
    \includegraphics[width=0.45\textwidth]{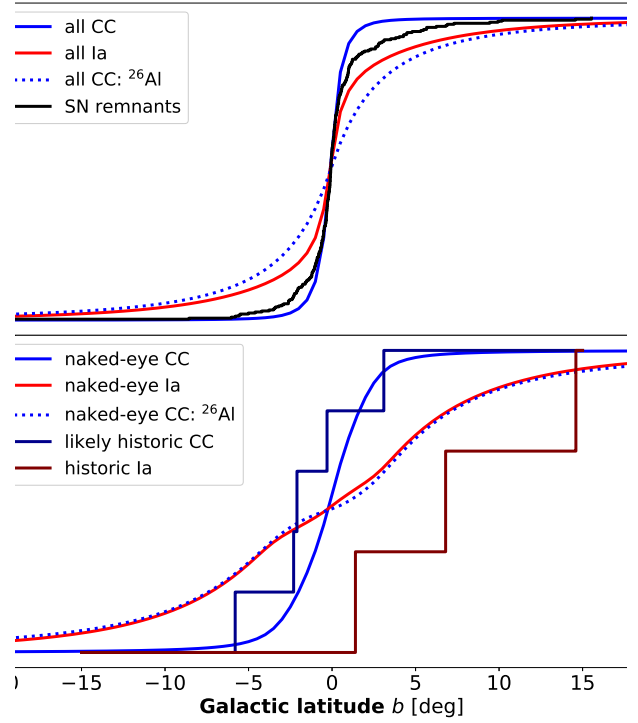}
    \caption{Cumulative distribution of supernova latitudes.  {\em Top:} intrinsic distributions
    compared to the \citet{Green2019} catalog of observed SNRs. {\em Bottom:} naked-eye distributions
    compare to historic supernovae.}
    \label{fig:lat-cume}
\end{figure}

\begin{figure}
    \centering
    \includegraphics[width=0.45\textwidth]{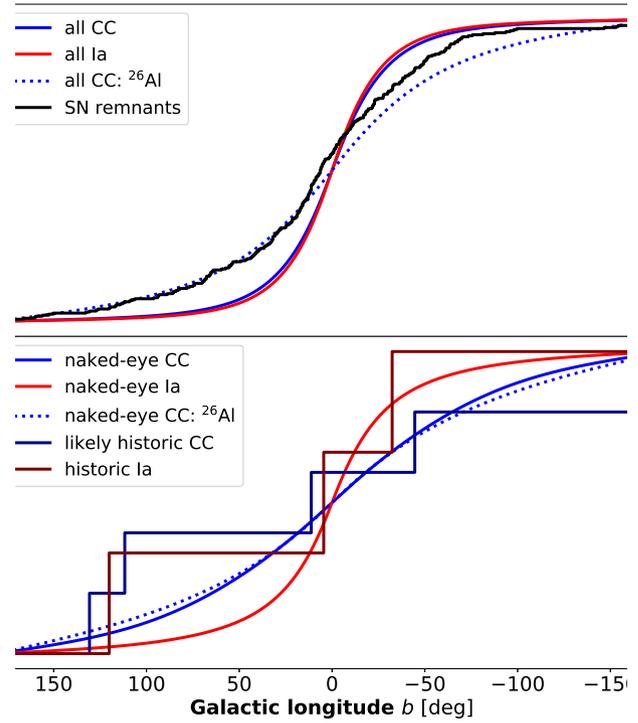}
    \caption{As in Fig.~\ref{fig:lat-cume} but for the cumulative distributions in longitude.}
    \label{fig:long-cume}
\end{figure}

Comparing the observed remnants and the model prediction
reveals areas of both agreement and disagreement.  We see that the theory contours enclosing 68\% of the total probability (darker shades of blue and red) are in rather good agreement with the remnant 68\% region (solid black line).
All three have a similar longitude extent.  In latitude,
the observed remnant width is larger than the model CCSN
and smaller than the model SNIa, as we would expect given that the
remnants sum both.  On the other hand, we note that the 95\% contours for the models (lighter blue and red shades) and the data (dashed gray line) are not particularly well-matched.  In particular,
the SNIa model predicts more high latitude events than are seen in remnants.  And the observed remnants in the outer quadrants have a thicker width than either model.  

The sky distribution of remnants
has some notable features
suggestive of incompleteness.
We see that there are
more high-latitude events
in the outer quadrants (i.e.,
$|\ell| > 90^\circ$) than in inner quadrants. As \citet{Green2015}
notes, this likely reflects
the lower sensitivity in the Galactic
center due to the higher background
of diffuse radio emission in that
region.
There is also a notable asymmetry in the counts
at the outer quadrants,
with higher counts in the second
quadrant than the third.  
This too was noted by \citep{Green2015},
who points out that
the second quadrant is in
Earth's northern hemisphere and thus
has enjoyed more and deeper study at radio
wavelengths.  
In summary, Fig.~\ref{fig:SNR-Sky} suggests
there remain supernova remnants remain
to be found at higher latitudes
in the inner quadrants, and in
the southerly third quadrant.

The placement of historical supernovae in
Fig.~\ref{fig:SNR-Sky} 
is revealing.
The map reveals the confirmed historical supernovae, 
marked as stars and labelled,
seem to be for the most part outliers
in the SN remnant distribution.
All five confident events lie outside the 68\% contour
of the remnant population, and both SN 1006 and SN 1054
lie outside the 95\% contour.
All historical supernovae are found high
latitudes compared to the 
rest of the distribution.
The two confirmed core-collapse
supernovae SN 1054 and SN 1181 both lie in the two
outer Galactic quadrants;
SN 1054 is also at a high latitude
and nearly at the Galactic anticenter.
Of the three confirmed historical
Type Ia SNe,
SN 1572 is in an outer quadrant.
SN 1006 is in an inner quadrant
but has the 
third highest $|b|$
of the 294 object sampled!
Indeed, of all
confident historical supernovae
only SN 1604 is in a region
at low latitude near the
Galactic center.

Of course, it is to be expected
that historical supernovae
do not follow the full
distribution of Galactic
supernovae because
naked-eye observations
and historical records
only capture the brightest
events.
To properly compare 
the historical SN events
with expectations,
we must account 
for their observability,
to which we now turn.

\subsection{The Distribution of Naked-Eye Events and Comparison With Historical Supernovae}

After taking into account the dust and its visible band extinction, we arrive at the sky plots in Figures \ref{fig:cc_w_ex} and \ref{fig:Ia_w_ex}. These plots describe the probability density
of a visible SN occurring in the sky.
The black contours enclose 68\% of the total probability,
while the gray contours enclose 95\%.  The addition of dust extinction shows that the Galactic center is no longer the highest probability,
due to the large optical depths to the Galactic center and beyond. Two regions slightly above and below the center of the Galaxy become the highest probability of observing a SN for both core-collapse and Type Ia 
explosions.  

Furthermore, for both core-collapse and Type Ia events, the peak in the region slightly north of the Galactic center is higher than its counterpart south of the center.
This asymmetry arises from the Sun's non-zero height above
the Galactic plane.  As noted in the previous section,
the source density is higher for sightlines at negative latitudes.
But the dust density is also slightly higher for
the same sightlines, and the effect of extinction dominates.

We have adopted a minimum apparent magnitude
$m_V^{\rm lim}=2$
for Figs.~ \ref{fig:cc_w_ex} and \ref{fig:Ia_w_ex}
and supernova magnitudes corresponding to the 
$\Delta t = 90$ day sustained luminosity 
of $(M_V^{\rm CC},M_V^{\rm Ia})=(-15,-14.7)$.
Since these luminosities are very similar, 
the differences between the plots almost entirely reflect
the difference in the underlying geometry, i.e., that
the SN Ia distribution has a thick-disk component.
We have explored the effect of choosing 
any one of these magnitudes to be brighter
or dimmer by $\pm 2  \ {\rm mag}$.
The changes are not dramatic,
but as expected, the distributions become 
somewhat broader in their extension in both 
latitude and longitude when made dimmer.

Figures  \ref{fig:cc_w_ex} and \ref{fig:Ia_w_ex} also show the
locations of the well-established historical supernovae.
Despite only having a few events, the comparison is striking,
as foreshadowed in Fig~\ref{fig:SNR-Sky}.
Core-collapse events appear in Fig.~\ref{fig:cc_w_ex},
where we see that the two confirmed historical discoveries, 
SN 1054 and SN 1181, 
both lie in regions of low probability.  
SN 1181 is at the edge of the 95\% contour, 
and SN 1054 is well outside of it.  Both events
are strongly offset in both latitude
and longitude from the predicted favored regions near the
Galactic center.  Indeed, the two supernovae
both lie in the outer (2nd and 3rd) Galactic quadrants,
with the Crab almost exactly at the anticenter!

Figure~\ref{fig:cc_w_ex}
also shows three potential historical CCSN events. 
Cas A has a location very similar to SN1181, lying in the 2nd quadrant and at a latitude
above the range our model favors, somewhat outside the 68\% contour but
well within the 95\% region.
By contrast, if SN386 is indeed
associated with the G11.2-0.3 remnant and pulsar, it lies very close to the Galactic center and is in the 
highest probability region 
of all of the events shown.
Finally, SN185 is less central
but still in the 4th quadrant 
and in a region of substantial probability--both this event
and SN 386 lie inside the 68\% contour.  Thus, if these two
candidate events could be confirmed, they will
partially relieve the mismatch between
the observed events and our
model predictions.

Turning to Type Ia events
in Figure \ref{fig:Ia_w_ex}, we find the situation is somewhat improved.
Once again, the historical discoveries avoid the regions of
highest probability around (but not precisely at) the Galactic center.  
But SN1604 and 1572 (Kepler and Tycho, respectively) both lie much closer to the high probability regions than the CCSN. 
Kepler lies at very central longitudes but is at a high
latitude compared to out peak region; it is inside the 68\% contour.   
Tycho is complementary in that it lies at low latitudes but at larger longitudes than our peak region--indeed, it is in the second Galactic quadrant.    
Finally, SN1006 is at moderate longitude compared to our predictions,
but lies at  $b\sim15$, a very high latitude relative to
our high-probability regions.  Indeed, SN 1006 has the 3rd highest $|b|$ of all of the supernova remnants in the \citet{Green2015} catalog.
Nonetheless, both SN 1006 and SN 1572 lie inside the 95\% region.


Thus we see that except for SN 1604,
the confident historical supernovae of both types
have sky locations outside of the regions our 
model favors.  Our calculations include
the effects of dust extinction,
so the discrepancies must reflect
one or more assumptions we have made.
Possible reasons for these mismatches are discussed below in \S \ref{sect:discuss}.

Fig.~\ref{fig:SN-summary} presents
a zoom-out summary of
our predictions and the historical data
for both supernova types.
Models shown are our fiducial 
model with $m_{\rm vis}=2 \ \rm mag$,
as in Figs.~\ref{fig:cc_w_ex} and \ref{fig:Ia_w_ex}.
The darker (lighter) blue and red regions enclose
68\% (95\%) of the total probability of seeing 
CCSN and SNIa respectively, as in as in Figs.~\ref{fig:cc_w_ex} and \ref{fig:Ia_w_ex}.
The distributions are thus broader than those in 
the intrinsic distribution seen in
Fig.~\ref{fig:SNR-Sky} because the visible supernovae
are closer.

Fig.~\ref{fig:SN-summary} also shows the celestial equator
and celestial poles.
As noted in \S \ref{sect:SNcats}
and in \citet{adams},
the best-studied historical supernova 
records are from civilizations in 
the northern hemisphere.  
These records thus are are less
sensitive to events in the southern sky,
and completely insensitive to
supernovae in the far south.
We note that SN 1006  at $\delta = -40^\circ$
is even further south than the Galactic center, yet was seen in the north, but it was
also the brightest SN known.  

It is therefore possible that 
there were additional historical SNe that were seen by only
southern observers.  
These events may have been recorded in some way.
For example, the southernmost part of the
Galactic plane is at $\ell = 302.9^\circ = -57.1^\circ$.
\footnote{The maximum excursion of the Galactic plane from the celestial equator has a declination $|\delta| = 90^\circ - b_{\rm NCP}=62.9^\circ$
at longitudes $\ell=(\ell_{\rm NCP},\ell_{\rm SCP})=(122.9^\circ,302.9^\circ)$,
where $(l_{\rm NCP},b_{\rm NCP})$ is the north celestial pole location in Galactic coordinates.}
This lies in the inner region of the plane--the 4th quadrant--and
has significant probability in both our CC and Ia models.

Furthermore, southern observers would have
had excellent views of SN 1006, and better
views of SN 1064; both would have been
much higher above the horizon
than for northern observers
and thus visible for longer, from
more locations, and with less air-mass.  
Furthermore, 
the proposed SN185 and 386 locations
lie in the southern sky,
and would be of particular
interest to find in historical records.
On the other hand, by unfortunate coincidence, SN 1181 and 1572 are both at longitudes
close to $\ell = 123^\circ$,
where the Galactic plane has its maximum northern excursion,
and both lie at $\delta \simeq +65^\circ$, making them two of the northernmost known supernova remnants.
Thus, these events would have been difficult or impossible to
be seen from all but the lowest southern latitudes.

We thus encourage efforts to 
recover and study historical astronomical
records from southern hemisphere.
These could provide new insights
into known or suspected historical supernovae.
Indeed, the known events
can also serve as ``calibration''
points in interpreting ancient records.
And it is possible that completely new 
events await (re)discovery
in southern archives.

\subsection{Marginalized Results:  Latitude and Longitude Distributions}

We can marginalize our 2-D all-sky distributions to obtain
1-D distributions of supernovae vs latitude or longitude.
The latitude distribution is given by
\begin{equation}
\frac{dP}{db} = \int \frac{dP}{d\Omega} \ \frac{d\Omega}{db}
= \frac{1}{\cos b} \int \frac{dP}{d\Omega} \ d\ell \ .
\end{equation}
This appears in Fig.~\ref{fig:lat-dis}
that shows both intrinsic and naked-eye cases
in our fiducial model and for the for 
the \al26-based CCSN distribution.
As expected, we see that for both intrinsic and naked-eye cases,
the fiducial CCSN
distribution is the narrowest,
followed by the SNIa then \al26 models.
This reflects the differences in scale heights.
We note that the SNIa distribution
averages thin and thick disk components corresponding
to the fiducial and \al26-based CCSN models.  Thus the
SNIa latitude distribution is intermediate.
Also as expected, we see that the naked-eye events have
a wider latitude distribution than their corresponding
intrinsic cases. This reflects the shorter distances
to the naked-eye events, leading to the scale height 
subtending a larger angle.

\begin{table}    \centering
    \caption{Central Intervals for Marginalized Distributions in Fiducial Models}
    \begin{tabular}{c|cc|cc}
    & \multicolumn{2}{c}{latitude $|b| \, \rm [deg]$}  & \multicolumn{2}{c}{longitude $|\ell| \, \rm [deg]$} \\
    Model & 68\%CL & 95\%CL &
    68\%CL & 95\%CL  \\
    \hline
    Intrinsic CCSN & 0.7 & 2.4 & 29 & 88\\
    Intrinsic SNIa  & 2.5 & 14 & 27 & 79 \\
    Intrinsic CCSN:\al26  & 4 & 17 & 64 & 152 \\
    \hline
    Naked-Eye CCSN  & 2 & 6 & 29 & 88 \\
    Naked-Eye SNIa & 8 & 24 & 35 & 118  \\
    Naked-Eye CCSNe:\al26 & 9 & 29 & 87 & 164 \\
    \hline\hline
    \end{tabular}
    \label{tab:cume}
\end{table}

The longitude distribution 
is given by
\begin{equation}
\frac{dP}{d\ell} = \int \frac{dP}{d\Omega} \ \frac{d\Omega}{d\ell}
=  \int \frac{dP}{d\Omega} \ \cos b \ db \
\end{equation}
and appears in Fig.~\ref{fig:long-dis}.  
Here the scale radius is the controlling factor.
Thus, the intrinsic CCSN and SNIa distributions are very similar, reflecting the similar scale radii, while
the intrinsic \al26 distribution is broader, 
For the case of the naked-eye events, the extinction towards the
Galactic center down-weights this region strongly for 
the CCSN, leading to a broader longitude distribution.  
This effect is much less pronounced for the other cases
where the scale height is larger.

Finally, we can integrate the 1-D results to
obtain cumulative distributions in both latitude
and longitude:
\begin{eqnarray}
P(<b) & = & \int_{-\pi/2}^b \frac{dP}{db} \ \cos b \ db \\
P(>\ell) & = & \int_{\ell}^{\pi} \frac{dP}{d\ell} \ d\ell
\end{eqnarray}
These are of particular interest because these
allow direct comparison with 
the cumulative distributions of
observed SNe.
These also give all-sky total probabilities of observable supernovae to be
$P_{\rm CC,tot} = 14\%$ and $P_{\rm Ia,tot} = 33\%$.

Fig.~\ref{fig:lat-cume} 
shows our predicted cumulative latitude distributions,
and Table \ref{tab:cume} quantifies the angular scales at which 68\% and 95\% of events are contained.
The top panel gives the results for our 
intrinsic models shown in the upper panel of Fig.~\ref{fig:lat-dis}.  To represent
observations, we show the cumulative distribution
of the \citet{Green2019} SNR catalog
tha combines CCSN and SNIa;
this is somewhat wider on the $|b|>0$ side,
perhaps reflecting the bias towards more extensive observations
from the terrestrial northern hemisphere.
We see that the observed SNR data lies between
the predictions of fiducial CCSN and SNIa models,
which is as expected given that the data
includes remnants from both SN types.
On the other hand, the very broad distribution
based on \al26 observations is in strikingly poor agreement
with the observer SNR population.  This suggests
that either \al26 has important sources other than SNe,
or it could be due to the Sun's embedding in local \al26 source 
\citep{Fujimoto2020a,Krause2020}; see discussion in \S\ref{sect:discuss}.

In the lower panel of Fig.~\ref{fig:lat-cume} we show the cumulative distribution of naked-eye supernovae, corresponding to the lower panel of Fig.~\ref{fig:lat-dis}.  We see that the narrow fiducial CCSN distribution gives sharp rise around $\ell=0$, while the wider SNIa and \al26-based distributions show a much more gradual rise.  We also show the cumulative distributions for confident and likely historic supernovae of each type.  We see that the historic CCSN events extend to larger $|b|$ than predicted in the model, particularly the first step that corresponds to SN 1054.  Turning to SNIa, it is striking that all three events are in the Galactic north, which leads to a strong mismatch with the predictions that are by construction north-south symmetric.  This aside, it is again clear from the plot and from Table \ref{tab:cume} that our model does not assign high probability to high-latitude events.
Finally we note that the \al26-based distribution is very similar to that of SNIa, reflecting the same large scale height value used in both.

The corresponding cumulative longitude distributions appear in Fig.~\ref{fig:long-cume} for the same models and observed populations
as in Fig.~\ref{fig:lat-cume}. 
In the top panel we show the intrinsic models. We see that the fiducial CCSN and SNIa cumulative
longitude distributions are very similar, reflecting the similar scale radii, while the \al26-based distribution is substantially broader.  Interestingly, the \citet{Green2019} SNR distribution is broader at positive latitudes than at negative latitudes, likely reflecting the bias towards more detections in the north.  Furthermore, the SNR curves are broader than both the fiducial models, and therefore the observed SNR slope near $\ell=0$ is shallower.  This could reflect the bias against SNR detection in the central regions of the Galactic plane due to crowding.
Intriguingly, the \al26-based distribution agrees better with the SNR data, providing a particularly good match for the positive latitude data.  Thus the observed SNR latitude and longitude distributions favor different models, though biases are likely to influence the comparison.

Finally, the bottom panel of \ref{fig:long-cume} shows the cumulative longitude distribution of the naked-eye supernovae.  Here the CCSN and \al26-based models give similar results and are much broader than the SNIa case.  The historical supernovae of both types depart significantly form all of the predictions.  As expected, this is driven by the events at large $\ell$ that are not predicted in any of the models.
We thus conclude that difficulty in explaining these outer-quadrant events is a quite general feature of our models.  This suggests that the solution will require we relax one or more of our basic assumptions, as we will discuss more below.

\subsection{Historic Events and Milky-Way Supernova Rates}

\begin{figure*}
    \centering
    \includegraphics[width = \textwidth]
    {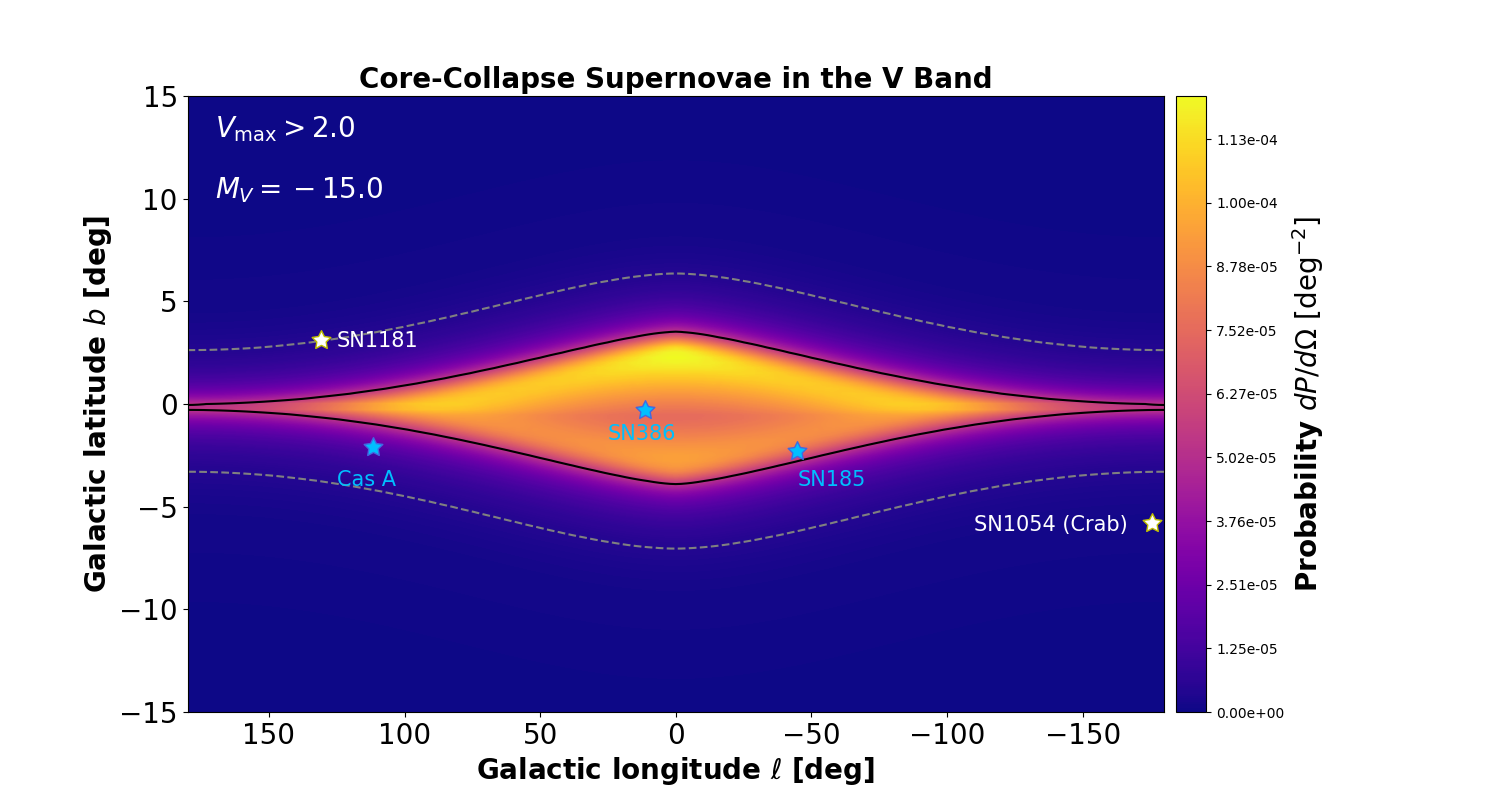}
    \caption{Sky plot of the probability density
    for core-collapse supernovae observed at Earth with apparent magnitude $m_ V \ge 2 \ {\rm mag}$.  The coordinates and stretched vertical axis are as in figs.~\ref{fig:cc_wo_ex} and \ref{fig:Ia_wo_ex}. 
    Effects of dust extinction is taken into account, and the supernova absolute magnitude is taken to be $M_V = -15$,
    representative of the 90 day
    sustained CCSN luminosities in Table \ref{tab:SNabsmag}.
    Confirmed historical CCSNe are shown in white stars, possible historical events are shown as shaded stars.
    The all-sky probability integrates to $P_{\rm CC,tot} = 14\%$ of
    all CCSN are visible to the naked eye;
    the black (gray) contour encloses 68\% (95\%) of this total probability.}
    \label{fig:cc_w_ex}
\end{figure*}


\begin{figure*}
    \centering
    \includegraphics[width = \textwidth]    {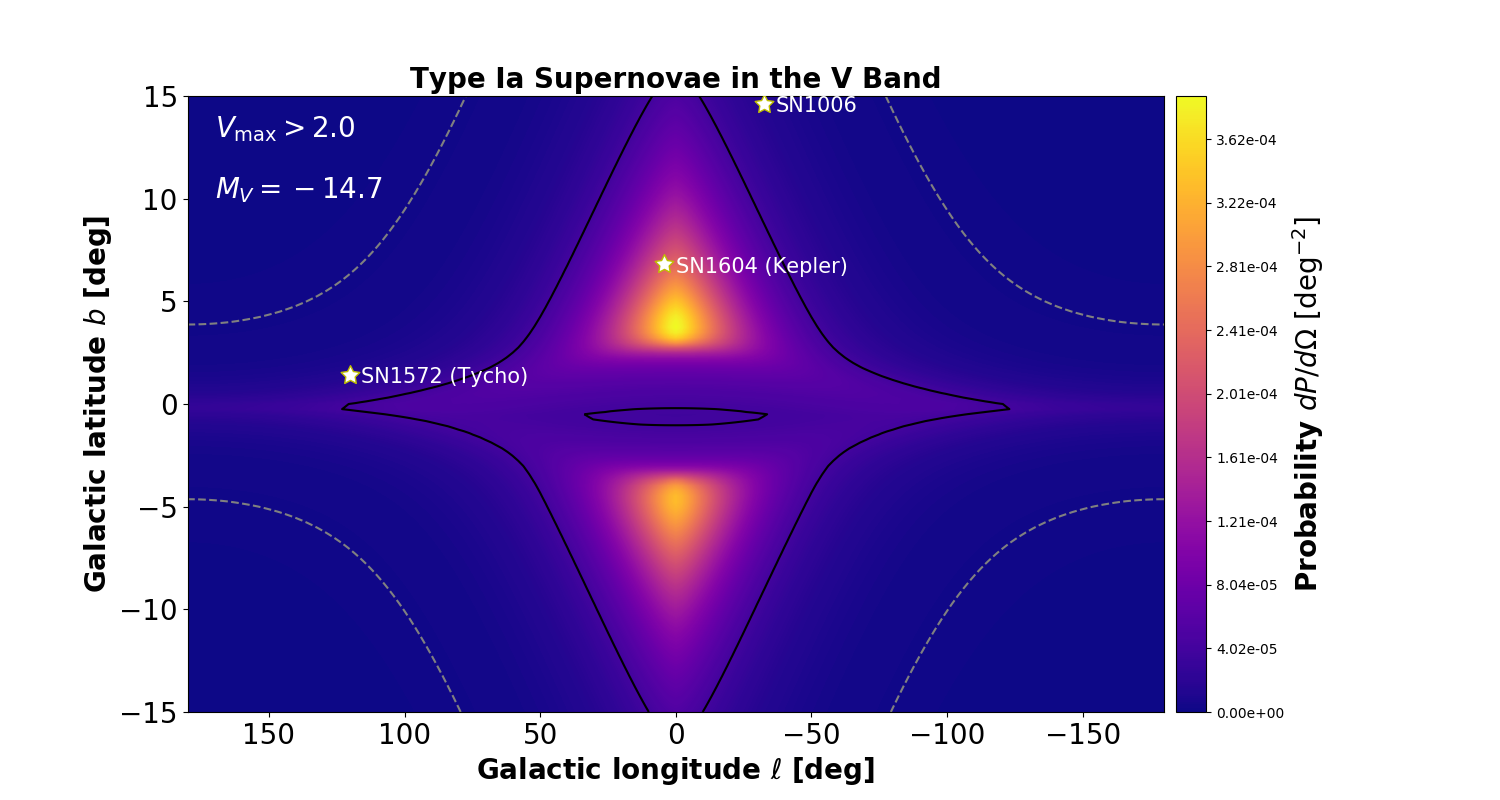}
    \caption{Events brighter than $m_V = 2$ as in Fig.~\ref{fig:cc_w_ex}, but for Type Ia supernovae, taken to have absolute magnitude $M_V = -14.7$ corresponding to the the 90 day sustained luminosity in Table \ref{tab:SNabsmag}. The all-sky probability integrates to $P_{\rm Ia,tot} = 33\%$ of all SNIa are visible to the naked eye;
    Tte black (gray) contour encloses 68\% (95\%) of this total probability,}
    \label{fig:Ia_w_ex}
\end{figure*}

\begin{figure*}
    \centering
    \includegraphics[width=\textwidth]{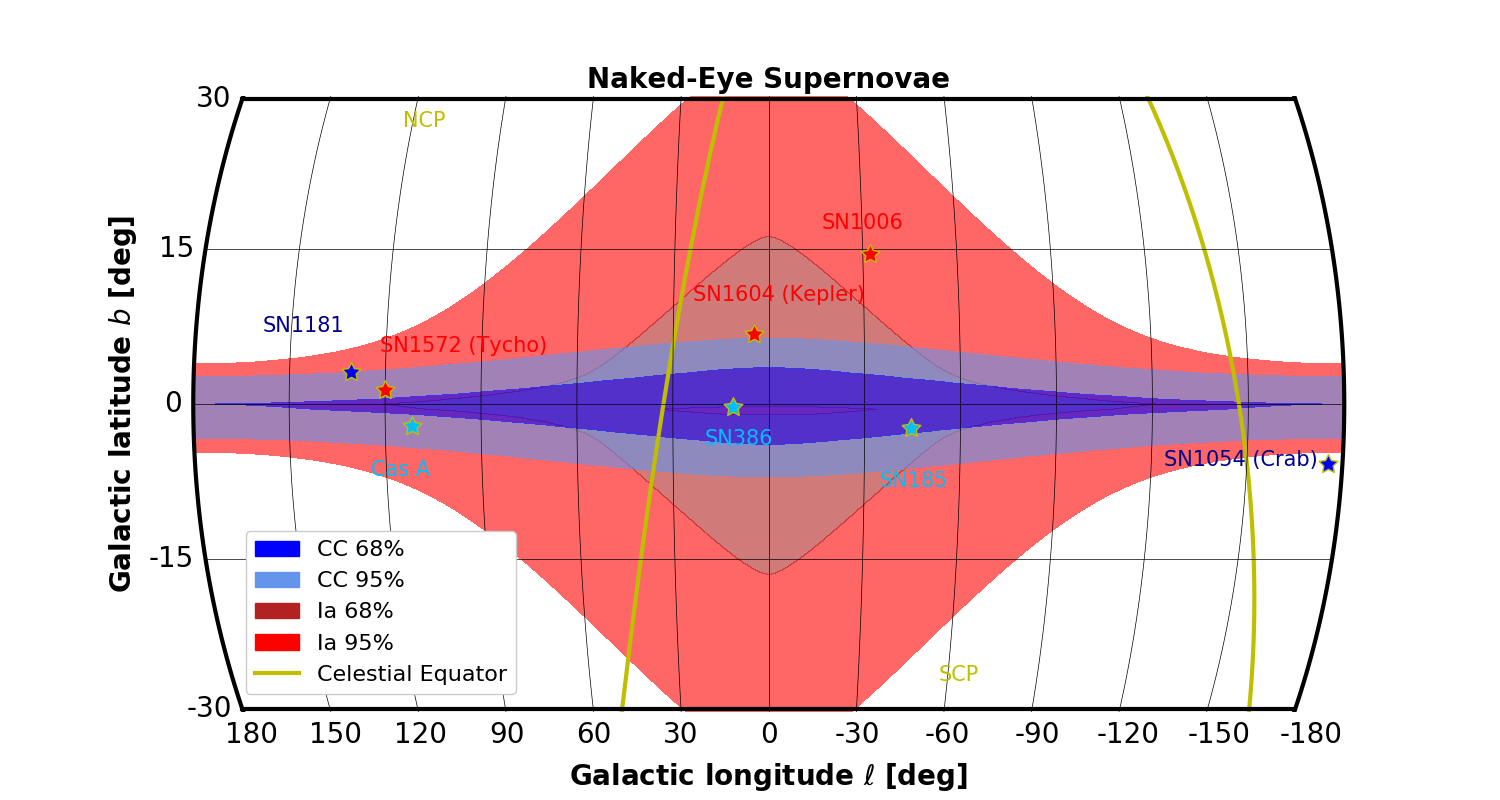}
    \caption{Zoom-out map in Galactic coordinates of naked-eye supernova probability density
    similar to  Fig.~\ref{fig:SNR-Sky}.
    Visible probability density
    show for CCSN as in Fig.~\ref{fig:cc_w_ex} and Type Ia events as in Fig.~\ref{fig:Ia_w_ex}.  
    Historical supernovae are
    labeled as in Fig.~\ref{fig:SNR-Sky};
    we see that the confirmed historical events avoid the far southern sky, likely reflecting the lack of available information from civilizations in the southern hemisphere.}
    \label{fig:SN-summary}
\end{figure*}

\begin{figure*}
    \centering
    \includegraphics[width = \textwidth]{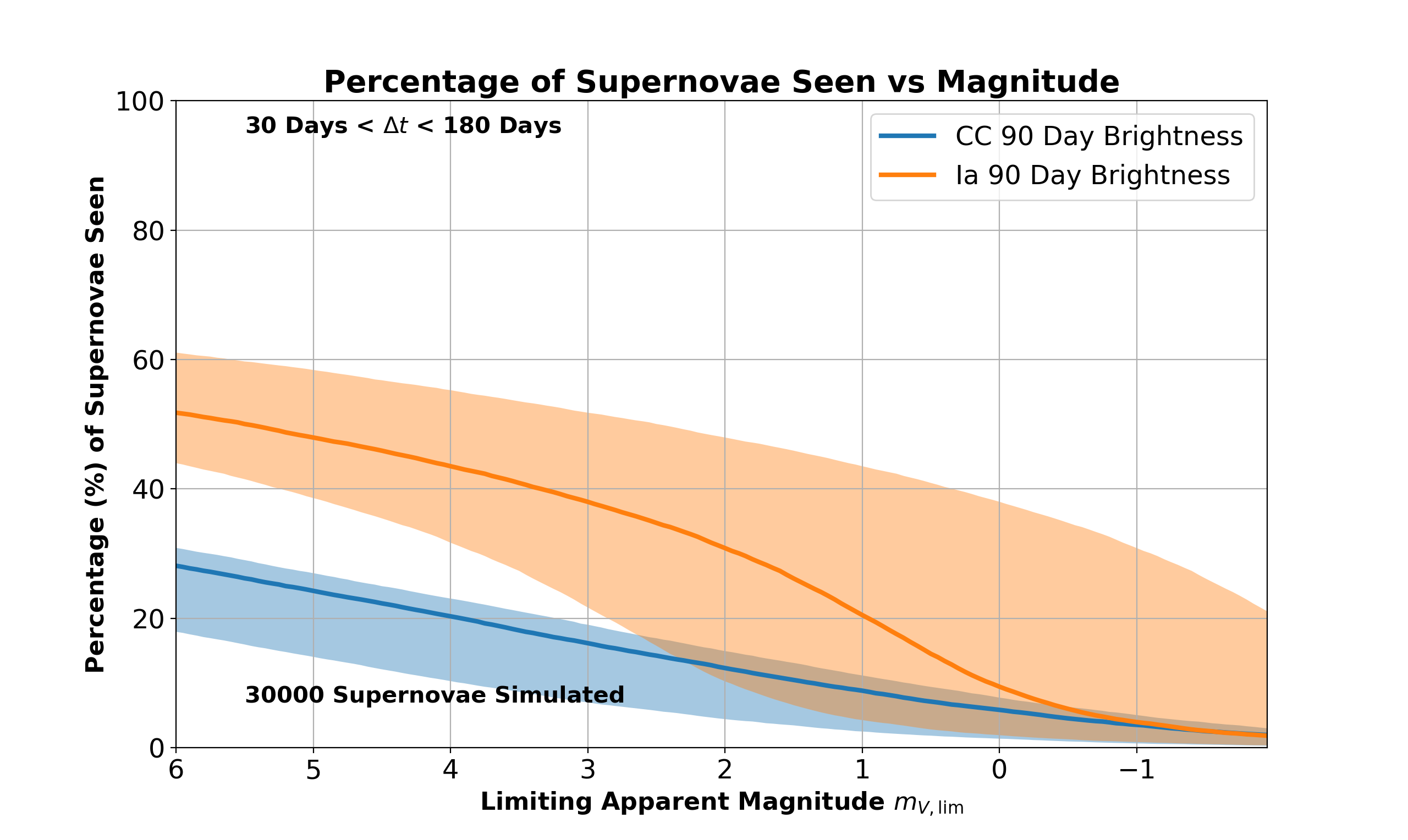}
    \caption{Plot of percentage of both types of supernovae seen for limiting magnitudes $m_{V,\rm lim}$ set to different sustained luminosities $M_V(\Delta t)$.  Solid curves are for $\Delta t = 90 \ \rm days$,
    upper (lower) band is for$\Delta t$ = 30 (180) days.  Orange curves are for Type Ia events, and blue curves are for CCSN.}
    \label{fig:viz_vs_mag}
\end{figure*}

Our model predicts the fraction
of all events that are visible to 
the naked eye.  This corresponds to an integral of the sky density
as in eq.~(\ref{eq:Ptot}),
but the Monte Carlo approach is
particularly useful for this 
calculation.  For both CCSN and SNIa,
we create a realization
of 30,000 events,
and ask what fraction are visible
from the Sun's location,
as a function of the limiting magnitude
$m_{V,\rm lim}$.  Results
appear in Fig.~\ref{fig:viz_vs_mag},
with the solid curves using
the sustained luminosity of
$M_V(90 \, \rm days)$
as in Table \ref{tab:SNabsmag}.
At $m_{V,\rm lim}=2$
we find $(f_{\rm vis,CC},f_{\rm vis,Ia}) = (0.33,0.13)$,
in good agreement with our results from the direct integration method.
The upper and lower bands around each
curve correspond to the 
results for the sustained luminosity
at $\Delta t = 30 \ \rm days$
and 90 days, respectively.

Figure \ref{fig:viz_vs_mag}
shows that for both SN types,
the observability drops as we demand a brighter limiting magnitude; this is
as expected. The SNIa fraction is higher because more events occur out of the midplane.
Around $m_{V,\rm lim} \sim 1$, the SNIa probability converges down to the CCSN probability.  
This reflects the fact that such bright events must occur so nearby
that extinction becomes unimportant.
In this limit, the fraction just corresponds to the small probability of
having very nearby events around the solar location.
The bands around the curves illustrate that the results are quite sensitive to
the duration required for supernova discovery.  Type Ia events show particularly strong variations, reflecting their strong drop in luminosity at late times.  

\begin{figure}
    \includegraphics[width=0.5\textwidth]{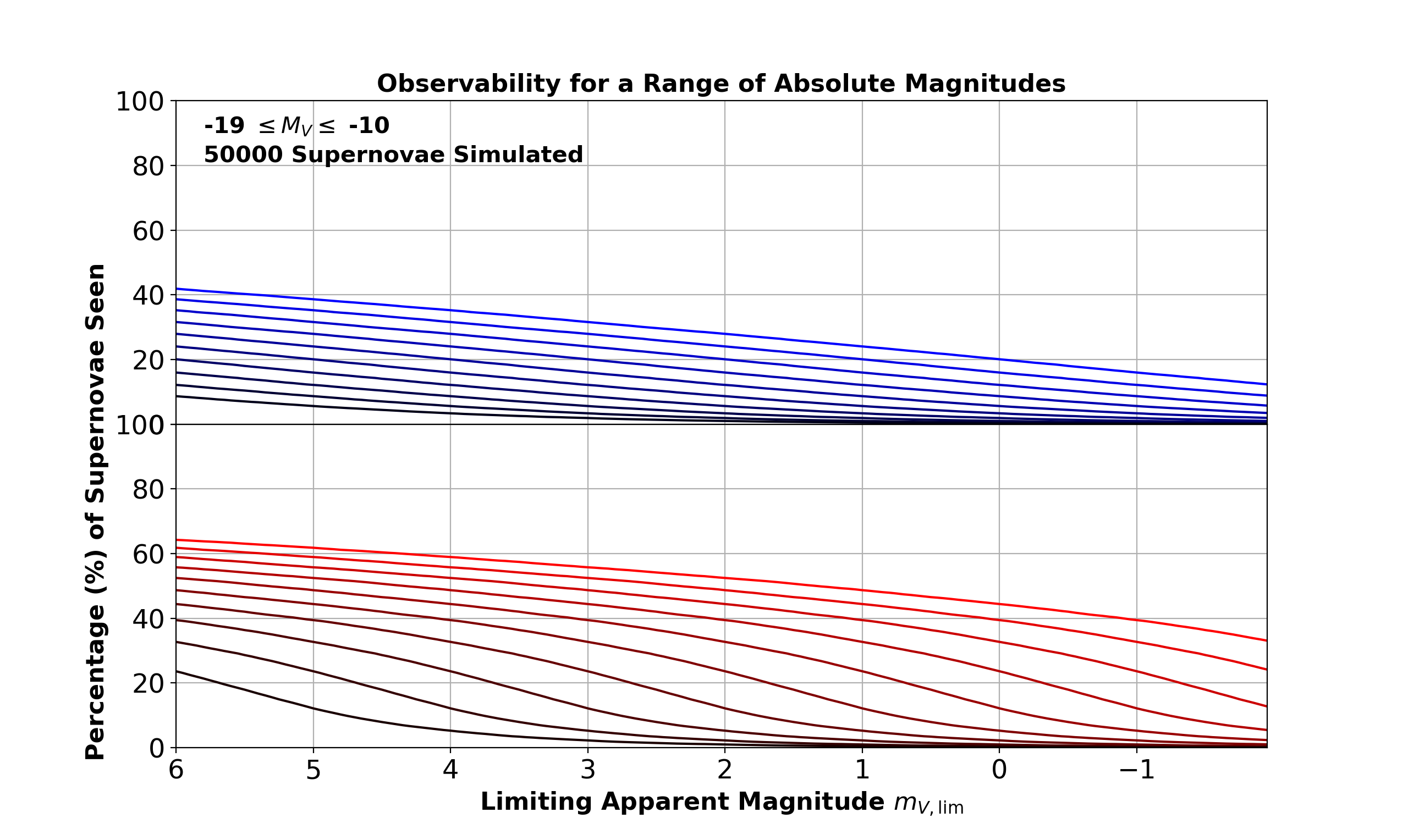}
    \caption{Fractions of supernovae visible as naked-eye events, plotted as a function of
    apparent magnitude as in Fig.~\ref{fig:viz_vs_mag};
    {\em top panel:} CCSN, {\em bottom panel:} SNIa.
    Shown for both supernova types,
    and for 10 values of absolute magnitude
    $M_V = (-19,-18,\ldots,-11,-10)$
    going from top to bottom.
    }
    \label{fig:viz_vs_mag2}
\end{figure}

Figure \ref{fig:viz_vs_mag2} 
also shows the fraction of supernova visibility
versus apparent magnitude,
now for a wide range of supernova absolute magnitudes.
This allows for estimation of the observability 
at any point along the light curve.  
For example, from Table \ref{tab:SNabsmag} we see that Type II-P supernovae
have $M_V = -17.0 \rm mag$ at peak but a sustained luminosity of 
$M_V(180 \ \rm days) = -13.9 \ \rm mag$. From 
the top panel of Fig.~\ref{fig:viz_vs_mag2} we see that at $m_{V,\rm lim}^{\rm SN} =2 \ \rm mag$, these magnitudes correspond to detection probabilities of 
about 20\% and 6\% respectively.  This illustrates the 
importance of the timescale needed for supernova discovery
and confirmation.

We can use historical supernova observations, along with
our observable fractions, to estimate the underlying
global Milky Way supernova rates.  For supernova type $i$, the expected number
$N_{\rm obs}$ of events observed over a duration $\Delta t_{\rm obs}$
\begin{equation}
    N_{\rm obs} = f_{\rm vis} f_{\rm night} {\cal R}_i \, \Delta t_{\rm obs}
\end{equation}
Note that any given region of the sky is not observable for 6 months because it is up during the day.  To account for this
we assign a fraction $f_{\rm night} \approx 1/2$ of
events that occur at night.
Thus we can estimate the global rate as
\begin{equation}
    R_i = \frac{N_{i,\rm obs}}{f_{{\rm vis},i} f_{\rm night} \, \Delta t_{\rm obs}}
\end{equation}
Of course the uncertainties are considerable due to the small
number of events, and so this exercise is best seen as a consistency
check of our model.  Using the historical data $(N_{\rm obs,Ia},N_{\rm obs,II}) = (3,2)$ over $\Delta t = 1 \ \rm kyr$, we estimate
\begin{eqnarray}
{\cal R}_{\rm Ia} & = & 1.8_{-1.2}^{+1.4} \ \pfrac{0.33}{f_{\rm vis,Ia}} \ {\rm events/century} \\
{\cal R}_{\rm CC} & = & 3.1_{-1.9}^{+3.5} \ \pfrac{0.15}{f_{\rm vis,CC}} \ {\rm events/century}
\end{eqnarray}
where the errors account only for Poisson counting statistics
\citep{Feldman1998}.
These are in good agreement with estimates
in Table \ref{tab:sn_types}
from the compilation in \citet{adams};
which performed a similar (but not identical) analysis using
the same historical supernovae.
Our result is thus also consistent within errors 
with the \citet{Rozwadowska2021} global mean of 
supernova rate estimators, though at the high end
of their range, as is typical of
estimates base on rates inferred from the
solar neighborhood \citep[e.g.,][]{vandenBergh1990,Dragicevich1999,Reed2005}.


Finally, Fig.~\ref{fig:viz_vs_mag}
invites us to consider the implications
for the next Galactic supernova
for naked-eye stargazers.  
Based on 
the intrinsic rates in Fig.~\ref{fig:SNTypes},
a CCSN is the most likely next event.  We see
that even at nominal human detection limit $m_{V,\rm lim} \simeq 6$,
there is only about a 1 in 3 chance 
that such an event will be visible
to the naked eye.  A brighter
event is even less likely.
For the next SNIa, there is more hope.
The odds are about 1 in 2 that 
it will be at the threshold of human
vision.  And the odds remain as high as about 1 in 3 that it will attain a
sustained apparent brightness of at least $m_V=2$
for 90 days,
which would provide a memorable
sight.

\section{Discussion}
\label{sect:discuss}

Despite the relatively small numbers of historical supernovae,
their observed sky positions and time history give
surprising insights into our model for naked-eye events.
We have shown that historical (naked-eye) supernova sky positions largely avoid the peaks of the probability distributions we predict.  This is acutely striking in the of core-collapse events, where
SN 1181 and the Crab that both lie in the outer Galactic quadrants near the anticenter, both in regions that are low-probability  for {\em any} of the axisymmetric exponential disk models we have considered.  On the other hand putative locations of SN 185 and especially SN 386 are in regions
our model favors.  For historical Type Ia events the match is
somewhat better, yet here too there are no events in the predicted highest probability regions.

This mismatch between
predictions and historical observations is likely real and implies that our model is incomplete or incorrect.  The patterns our model fails to capture are (1) an excess of events away from the central regions of the Galaxy, (2) an excess of events at relatively high latitude, and (3) a deficit of events near the Galactic center.
These patterns suggest several possible solutions.

One possibility is that our assumption of axisymmetry has omitted important structures in the stellar or dust distributions. 
In this paper, we adopt an axisymmetric model of the Galaxy. This ignores the existence of spiral arms and assumes the disk is smooth. 
But observations of extragalactic supernovae confirm 
that core-collapse events are correlated with spiral arms,
and that Type Ia events are as well, albeit more weakly \citep[e.g.,][]{Aramyan2016}. 
Models for the Galactic pulsar distribution also
favor compact object birth in spiral arms \citep{Faucher-Giguere2006}.
Indeed, \citet{Dragicevich1999} suggest that supernova
concentration in spiral arms leads to an anomalously high
supernova rate near the Sun's location in the Galaxy,
and thus a higher Galactic rate using historical data
than from other methods.  
Conversely, \citet{Li1991} studied the observed Milky-Way supernova remnant distribution available at the time, 
and did not find a statistically significant association
with spiral arms, though they suggest the observational
bias might obscure such a correlation.  
Finally, we note that \citet{Fujimoto2020b} study the likelihood of stars located in regions
near supernova explosions and note that these are most likely just outside spiral arms.

The historical events themselves seem to show an affinity for spiral
arms.
The Crab is found in the Perseus spiral arm, near the point in arm that is closest to the Sun. SN 1181 also lies in the Perseus arm, and Cas A may as well.  SN 185 and SN 1006 may lie in or near the Sagittarius arm.
Clearly, models with spiral structure merit investigation, and
we plan to do so in future work.

It is striking that our global supernova
rate estimates are in reasonable agreement 
with other methods, yet our sky distributions are generally 
poor matches
for the locations of historical supernovae.
This could be a coincidence, or it could suggest
that our model reasonably captures the the statistics of supernova obscuration
while not accounting for local departures from axisymmetry.


The lack of well-studied historical records from the southern
hemisphere raises the question of whether there have been historic
events in the south that are absent from northern records.
Indeed, Fig.~\ref{fig:viz_vs_mag}
shows that our predicted CCSN
and SNIa peaks are in the southern hemisphere.
Mitigating against this is the record of the confirmed and suspected historical supernovae.
Of the five confident events, two were seen in the 
southern sky:  SN 1604 and SN 1006.
Notably, SN 1006 at $-42^\circ$
is substantially further south than the 
Galactic center at $-29^\circ$.
If we identify SN 185 with RCW 86, 
this would be discovery at $-62.5^\circ$,
near the southernmost excursion of the Galactic plane.
This seemingly demonstrates that at least some northern
historical observers could have seen
supernovae in and around the Galactic center.

Thus it is at least possible that there may be a true deficit of
reported supernovae in this region where our predictions peak.
A search for records in southern hemisphere civilizations \citep[e.g.,][]{Hamacher2014}
would be invaluable to confirm this, to cover regions
in the Galactic plane inaccessible from the north,
and to potentially gain new information on known historical events.

Another effect we have ignored is that that stars are generally formed in clusters, and CCSN progenitors in particular are formed in OB associations.  
Our model ignores spatial clustering, implicitly assuming that supernova locations are uncorrelated.  
To include this effect, one could view our probability
distribution as describing the locations of {\em clusters} rather than individual stars.  Then as long as the 
statistics of stars within clusters does not vary
across the Galaxy, then our predictions are unchanged.
We also note that the clustered birth of massive stars would
also lead to correlations of explosions in {\em time}
as well as space, but on timescales $\ga 10^6 \ \rm yr$
associated with star formation.  Thus we do not
expect multiple events from the same cluster in 
the historic record.


Finally, we note that our formalism for supernova probability distributions invites other applications.
One natural extension is to consider
very nearby supernovae  ($\la 100 \ \rm pc$)
that might deposit debris on Earth such as
radioactive ${}^{60}{\rm Fe}$, which
has indeed been detected in terrestrial and lunar
samples from an event $\sim 3$ Myr ago, 
as reviewed in \citet{Fields2019}.
\citet{Sovgut2020}
adapt our formalism to find the probability of
nearby events as a function of distance.
This updates earlier work going back to
\citet{Shklovsky1968}, and including
the lively \citet{Hartmann2002}
study of supernova and gamma-ray burst ``disturbance ecology,'' which 
takes into account the effects of the vertical profiles of the events themselves as well as Galactic dust.

\section{Conclusions}
\label{sect:conclude}

We summarize our results as follows:
\begin{itemize}
    \item 
    We have developed expressions to calculate the probability distribution on the sky for supernovae of different types, allowing for dust effects and an observer flux limit. We implement this formalism in the case of axisymmetric models of the Milky Way SNe and dust spatial distributions. 
    Our fiducial model is that of \citet{adams}, itself based on the
    TRILEGAL model for the thin and thick disks. 
    
    \item We present the intrinsic probability distributions on the sky for both CCSN and SNIa, ignoring the effects of dust obscuration and optical flux limit.
    As expected, for both supernova types, the probable regions are along the Galactic plane and concentrated in the central Galactic quadrants, with peaks near the Galactic center.
    The thick-disk component of SNIa has a larger scale height and slightly smaller scale radius than the thin-disk that hosts CCSN,
    lead to a SNIa distribution that stretches to larger latitudes but a slightly more narrow longitude range.
    
    \item 
    To model naked-eye events that comprise the historical SNe, we introduce a ``sustained luminosity'' observability criterion
    to account for the necessity of repeated observations to confirm supernovae as worthy of historical note. Table \ref{tab:SNabsmag} shows how the various SN types dim over time, and that the types that are initially brighter may end up becoming dimmer than other types on the same timescale.
    
    \item We present sky maps of CCSN and SNIa probabilities.  Our models predict the 
    intrinsic, underlying distribution of both
    to most probable near the Galactic center,
    and both are found at long the Galactic plane,
     CCSN at low latitudes and SNIa extending to higher
     latitudes.  For naked-eye events in the presence
     of dust extinction, the distributions still peak
     near the Galactic center and at relatively low latitudes.
     
     \item We compare out models with confirmed
     and likely historic supernovae.  The  
     confirmed events all {\em avoid} the high-probability
     regions of our fiducial models for both CCSN and SNIa;
     indeed the Crab is at high latitudes nearly at the
     anticenter. We conjecture that spiral arm effects
     can be important here and we are investigating this.
     
     \item
     Using our model we calculate the fraction of
     Galactic events visible to the naked eye, and use
     historical events to estimate Galactic rates,
     finding good agreement with other methods.
     
          \item
     As a contrast to our fiducial model based on stellar disk populations, we present results
     for supernovae following the much more diffuse spatial distribution inferred from \al26 sky maps.
     We find that these sky maps are broader than our fiducial results in   both   and longitude; while the longitude distribution is 
     broadly consistent with the observed supernova remnant,
     the predicted latitude distribution is substantially broader.
     Also, even in this case, the highest latitude historic events
     are not well explained.

    \item Assuming a limiting visible magnitude of 2 over $\Delta t = 90$ days, we evaluate the chances of naked-eye visibility of
    the next Galactic supernova.  At the $m_V = 6$ limit, we find about a 33\% chance of seeing the next CCSN and about a 50\% chance of seeing the next SNIa.
\end{itemize}

There remains ample opportunity for future work.
We plan to relax the
assumption of axisymmetry, examining
the effect of spiral arms in the predicted event
maps. 
We encourage a search for indications of supernovae
and other transients in historical records from the southern
hemisphere.  
Finally, we look forward to deep surveys of the
southern sky in the radio
by the Square Kilometer Array and
in high-energy gamma rays by the Cerenkov 
Telescope Array; this
may detect the expansion of Galactic supernova
remnants, and reveal new remnants, perhaps including
events within the historical past \citep{Ingallinera2017}.





\section*{Acknowledgements}
It is a pleasure to acknowledge many useful discussions with Ashvini Krishnan, Danny Milisavljevic, Paul Ricker, Danylo Sovgut, Alexandra Trauth, and Xilu Wang.
BDF gratefully acknowledges fruitful discussions with all of the participants of the ``Historical Supernovae, Novae, and Other Transients'' workshop held in Oct.~2019, and the Lorentz Center at the University of Leiden for their hospitality in hosting this event.
An earlier version of this paper was submitted by JWH as a Senior Thesis
to the Department of Astronomy at the University of Illinois.

This work use ADS extensively, and used the Python packages
Matplotlib \citep{Hunter2007},
NumPy \citep{Oliphant2006},
SciPy \citep{Jones2001},
and AstroPy \citep{AstroPy2013}.





\begin{thebibliography}{9}

\bibitem[\protect\citeauthoryear{Abbott et al.}{2020}]{Abbott2020} Abbott B.~P., Abbott R., Abbott T.~D., Abraham S., Acernese F., Ackley K., Adams C., et al., 2020, PhRvD, 101, 084002

\bibitem[\protect\citeauthoryear{Adams, et al.}{2013}]{adams} Adams S.~M., Kochanek C.~S., Beacom J.~F., Vagins M.~R., Stanek K.~Z., 2013, ApJ, 778, 164

\bibitem[\protect\citeauthoryear{Al Kharusi et al.}{2020}]{SNEWS2020} Al Kharusi S., BenZvi S.~Y., Bobowski J.~S., Brdar V., Brunner T., Caden E., Clark M., et al., 2020, arXiv, arXiv:2011.00035

\bibitem[\protect\citeauthoryear{Antonioli, et al.}{2004}]{Antonioli2004} Antonioli P., et al., 2004, NJPh, 6, 114

\bibitem[\protect\citeauthoryear{Aramyan, et al.}{2016}]{Aramyan2016} Aramyan L.~S., et al., 2016, MNRAS, 459, 3130

\bibitem[\protect\citeauthoryear{Astropy Collaboration et al.}{2013}]{AstroPy2013} Astropy Collaboration, Robitaille T.~P., Tollerud E.~J., Greenfield P., Droettboom M., Bray E., Aldcroft T., et al., 2013, A\&A, 558, A33. doi:10.1051/0004-6361/201322068


\bibitem[\protect\citeauthoryear{Bartunov, Tsvetkov, \& Pavlyuk}{2007}]{Bartunov2007} Bartunov O.~S., Tsvetkov D.~Y., Pavlyuk N.~N., 2007, HiA, 14, 316


\bibitem[\protect\citeauthoryear{Bianco et al.}{2014}]{Bianco2014} Bianco F.~B., Modjaz M., Hicken M., Friedman A., Kirshner R.~P., Bloom J.~S., Challis P., et al., 2014, ApJS, 213, 19

\bibitem[\protect\citeauthoryear{Bortle}{2001}]{Bortle2001} Bortle J.~E., 2001, Sky and Telescope, 101, 126



\bibitem[\protect\citeauthoryear{Brown et al.}{2014}]{Brown2014} Brown P.~J., Breeveld A.~A., Holland S., Kuin P., Pritchard T., 2014, Ap\&SS, 354, 89


\bibitem[\protect\citeauthoryear{Cinzano, Falchi, \& Elvidge}{2001}]{Cinzano2001} Cinzano P., Falchi F., Elvidge C.~D., 2001, MNRAS, 323, 34


\bibitem[\protect\citeauthoryear{Clark \& Stephenson}{1977}]{Clark1977} Clark D.~H., Stephenson F.~R., 1977, The Historical Supernovae, Pergamon Press, Oxford

\bibitem[\protect\citeauthoryear{Clark}{1990}]{Clark1990} Clark R.~N., 1990, Visual Astronomy of the Deep Sky, Cambridge:  Cambridge University Press

\bibitem[\protect\citeauthoryear{Dall'Ora et al.}{2014}]{DallOra2014} Dall'Ora M., Botticella M.~T., Pumo M.~L., Zampieri L., Tomasella L., Pignata G., Bayless A.~J., et al., 2014, ApJ, 787, 139


\bibitem[\protect\citeauthoryear{de Jaeger et al.}{2019}]{deJaeger2019} de Jaeger T., Zheng W., Stahl B.~E., Filippenko A.~V., Brink T.~G., Bigley A., Blanchard K., et al., 2019, MNRAS, 490, 2799

\bibitem[\protect\citeauthoryear{Doherty et al.}{2014}]{Doherty2014} Doherty C.~L., Gil-Pons P., Lau H.~H.~B., Lattanzio J.~C., Siess L., 2014, MNRAS, 437, 195. doi:10.1093/mnras/stt1877

\bibitem[\protect\citeauthoryear{Dragicevich, Blair \& Burman}{1999}]{Dragicevich1999} Dragicevich P.~M., Blair D.~G., Burman R.~R., 1999, MNRAS, 302, 693

\bibitem[\protect\citeauthoryear{DUNE collaboration et al.}{2020}]{DUNE2020} DUNE collaboration, Abi B., Acciarri R., Acero M.~A., Adamov G., Adams D., Adinolfi M., et al., 2020, arXiv, arXiv:2008.06647

\bibitem[\protect\citeauthoryear{Faucher-Gigu{\`e}re \& Kaspi}{2006}]{Faucher-Giguere2006} Faucher-Gigu{\`e}re C.-A., Kaspi V.~M., 2006, ApJ, 643, 332


\bibitem[\protect\citeauthoryear{Feldman \& Cousins}{1998}]{Feldman1998} Feldman G.~J., Cousins R.~D., 1998, PhRvD, 57, 3873

\bibitem[\protect\citeauthoryear{Fesen et al.}{2006}]{Fesen2006} Fesen R.~A., Hammell M.~C., Morse J., Chevalier R.~A., Borkowski K.~J., Dopita M.~A., Gerardy C.~L., et al., 2006, ApJ, 645, 283

\bibitem[\protect\citeauthoryear{Fields et al.}{2019}]{Fields2019} Fields B., Ellis J.~R., Binns W.~R., Breitschwerdt D., deNolfo G.~A., Diehl R., Dwarkadas V.~V., et al., 2019, BAAS, 51, 410

\bibitem[\protect\citeauthoryear{Fujimoto et al.}{2020}]{Fujimoto2020b} Fujimoto Y., Krumholz M.~R., Inutsuka S.-. ichiro ., Boss A.~P., Nittler L.~R., 2020, MNRAS, 498, 5532. doi:10.1093/mnras/staa2778

\bibitem[\protect\citeauthoryear{Fujimoto, Krumholz, \& Inutsuka}{2020}]{Fujimoto2020a} Fujimoto Y., Krumholz M.~R., Inutsuka S.-. ichiro ., 2020, MNRAS, 497, 2442. doi:10.1093/mnras/staa2125


\bibitem[\protect\citeauthoryear{Girardi, et al.}{2005}]{trilegal} Girardi L., Groenewegen M.~A.~T., Hatziminaoglou E., da Costa L., 2005, A\&A, 436, 895


\bibitem[\protect\citeauthoryear{Green}{2015}]{Green2015} Green D.~A., 2015, MNRAS, 454, 1517

\bibitem[\protect\citeauthoryear{Green \& Stephenson}{2017}]{Green2017} Green D.~A., 2017, in Handbook of Supernovae, A.~W.~Alsabti and P.~Murdin eds, Springer, p 37


\bibitem[\protect\citeauthoryear{Green}{2019}]{Green2019} Green D.~A., 2019, JApA, 40, 36

\bibitem[\protect\citeauthoryear{Green \& Gull}{1984}]{g1903_2} Green D.~A., Gull S.~F., 1984, Nature, 312, 527

\bibitem[\protect\citeauthoryear{Green \& Stephenson}{2017}]{Green2017a} Green D.~A., Stephenson F.~R., 2017, in Handbook of Supernovae, A.~W.~Alsabti and P.~Murdin eds, Springer, p 179

\bibitem[\protect\citeauthoryear{Green et al.}{2014}]{GreenG2014} Green G.~M., Schlafly E.~F., Finkbeiner D.~P., Juri{\'c} M., Rix H.-W., Burgett W., Chambers K.~C., et al., 2014, ApJ, 783, 114. doi:10.1088/0004-637X/783/2/114

\bibitem[\protect\citeauthoryear{Green et al.}{2019}]{GreenG2019} Green G.~M., Schlafly E., Zucker C., Speagle J.~S., Finkbeiner D., 2019, ApJ, 887, 93. doi:10.3847/1538-4357/ab5362


\bibitem[\protect\citeauthoryear{Guillochon et al.}{2017}]{Guillochon2017} Guillochon J., Parrent J., Kelley L.~Z., Margutti R., 2017, ApJ, 835, 64

\bibitem[\protect\citeauthoryear{Hamacher}{2014}]{Hamacher2014} Hamacher D.~W., 2014, JAHH, 17, 161

\bibitem[\protect\citeauthoryear{Hartmann, Kretschmer, \& Diehl}{2002}]{Hartmann2002} Hartmann D.~H., Kretschmer K., Diehl R., 2002, nuas.conf, 154


\bibitem[\protect\citeauthoryear{Hunter}{2007}]{Hunter2007} Hunter J.~D., 2007, CSE, 9, 90. doi:10.1109/MCSE.2007.55

\bibitem[\protect\citeauthoryear{Ikeda et al.}{2007}]{Ikeda2007} Ikeda M., Takeda A., Fukuda Y., Vagins M.~R., Abe K., Iida T., Ishihara K., et al., 2007, ApJ, 669, 519

\bibitem[\protect\citeauthoryear{Ingallinera et al.}{2017}]{Ingallinera2017} Ingallinera A., Trigilio C., Umana G., Leto P., Buemi C., Schillir{\`o} F., Bufano F., et al., 2017, IAUS, 331, 345. doi:10.1017/S1743921317004574

\bibitem[\protect\citeauthoryear{Jones et al.}{2001}]{Jones2001} Jones, E., et al., 2001, SciPy: Open Source Scientific Tools for Python.
Available at: \url{http://www.scipy.org/}

\bibitem[\protect\citeauthoryear{Katsuda}{2017}]{Katsuda2017} Katsuda, S., 2017, in Handbook of Supernovae, A.~W.~Alsabti and P.~Murdin eds, Springer, p 63

\bibitem[\protect\citeauthoryear{Kelly \& Kirshner}{2012}]{Kelly2012} Kelly P.~L., Kirshner R.~P., 2012, ApJ, 759, 107. doi:10.1088/0004-637X/759/2/107

\bibitem[\protect\citeauthoryear{Kelly, Kirshner, \& Pahre}{2008}]{Kelly2008} Kelly P.~L., Kirshner R.~P., Pahre M., 2008, ApJ, 687, 1201. doi:10.1086/591925


\bibitem[\protect\citeauthoryear{Kepler}{1606}]{kepler} Kepler J, 1606, De Stella Nova in Pede Serpentarii, Paul Sessius, Praha, Czech Republic, Available online: \url{https://www.univie.ac.at/hwastro/rare/1606_kepler.htm}

\bibitem[\protect\citeauthoryear{Kothes}{2013}]{Kothes2013} Kothes R., 2013, A\&A, 560, A18

\bibitem[\protect\citeauthoryear{Krause et al.}{2020}]{Krause2020} Krause M.~G.~H., Rodgers-Lee D., Dale J.~E., Diehl R., Kobayashi C., 2020, arXiv, arXiv:2011.08615


\bibitem[\protect\citeauthoryear{Krause et al.}{2008}]{Krause2008} Krause O., Tanaka M., Usuda T., Hattori T., Goto M., Birkmann S., Nomoto K., 2008, Natur, 456, 617

\bibitem[\protect\citeauthoryear{Kunkel et al.}{1987}]{Kunkel1987} Kunkel W., Madore B., Shelton I., Duhalde O., Bateson F.~M., Jones A., Moreno B., et al., 1987, IAUC, 4316


\bibitem[\protect\citeauthoryear{Li, et al.}{1991}]{Li1991} Li Z., Wheeler J.~C., Bash F.~N., Jefferys W.~H., 1991, ApJ, 378, 93

\bibitem[Li et al.(2011)]{Li2011} Li, W., Leaman, J., Chornock, R., et al.\ 2011, \mnras, 412, 1441

\bibitem[\protect\citeauthoryear{Luken, et al.}{2019}]{g1903_1} Luken K.~J., et al., 2019, MNRAS.tmp, 3072


\bibitem[\protect\citeauthoryear{Mauerhan et al.}{2013}]{Mauerhan2013} Mauerhan J.~C., Smith N., Silverman J.~M., Filippenko A.~V., Morgan A.~N., Cenko S.~B., Ganeshalingam M., et al., 2013, MNRAS, 431, 2599

\bibitem[\protect\citeauthoryear{Munari et al.}{2013}]{Munari2013} Munari U., Henden A., Belligoli R., Castellani F., Cherini G., Righetti G.~L., Vagnozzi A., 2013, NewA, 20, 30

\bibitem[\protect\citeauthoryear{Novoseltsev et al.}{2020}]{Novoseltsev2020} Novoseltsev Y.~F., Boliev M.~M., Dzaparova I.~M., Kochkarov M.~M., Kurenya A.~N., Novoseltseva R.~V., Petkov V.~B., et al., 2020, APh, 117, 102404

\bibitem[\protect\citeauthoryear{Oliphant}{2006}]{Oliphant2006} Oliphant, T., 2006, A Guide to NumPy, Trelgol Publishing, USA

\bibitem[\protect\citeauthoryear{Prentice et al.}{2018}]{Prentice2018} Prentice S.~J., Ashall C., Mazzali P.~A., Zhang J.-J., James P.~A., Wang X.-F., Vink{\'o} J., et al., 2018, MNRAS, 478, 4162

\bibitem[\protect\citeauthoryear{Reed}{2005}]{Reed2005} Reed B.~C., 2005, AJ, 130, 1652


\bibitem[\protect\citeauthoryear{Reynolds, et al.}{2008}]{g1903_3} Reynolds S.~P., Borkowski K.~J., Green D.~A., Hwang U., Harrus I., Petre R., 2008, ApJL, 680, L41

\bibitem[\protect\citeauthoryear{Rozwadowska, Vissani, \& Cappellaro}{2021}]{Rozwadowska2021} Rozwadowska K., Vissani F., Cappellaro E., 2021, NewA, 83, 101498


\bibitem[\protect\citeauthoryear{Sahu, Anupama, \& Chakradhari}{2013}]{Sahu2013} Sahu D.~K., Anupama G.~C., Chakradhari N.~K., 2013, MNRAS, 433, 2


\bibitem[\protect\citeauthoryear{Schlafly \& Finkbeiner}{2011}]{Schlafly2011} Schlafly E.~F., Finkbeiner D.~P., 2011, ApJ, 737, 103. doi:10.1088/0004-637X/737/2/103

\bibitem[\protect\citeauthoryear{Schlafly et al.}{2010}]{Schlafly2010} Schlafly E.~F., Finkbeiner D.~P., Schlegel D.~J., Juri{\'c} M., Ivezi{\'c} {\v{Z}}., Gibson R.~R., Knapp G.~R., et al., 2010, ApJ, 725, 1175. doi:10.1088/0004-637X/725/1/1175

\bibitem[\protect\citeauthoryear{Schlegel, Finkbeiner, \& Davis}{1998}]{Schlegel1998} Schlegel D.~J., Finkbeiner D.~P., Davis M., 1998, ApJ, 500, 525. doi:10.1086/305772


\bibitem[\protect\citeauthoryear{Scholberg}{2012}]{Scholberg2012} Scholberg K., 2012, ARNPS, 62, 81

\bibitem[\protect\citeauthoryear{Shivvers, et al.}{2017}]{Shivvers2017} Shivvers I., et al., 2017, MNRAS, 471, 4381

\bibitem[\protect\citeauthoryear{Shklovsky}{1968}]{Shklovsky1968} Shklovsky I.~S., 1968, 
Supernovae, Interscience Monographs and Texts in Physics and Astronomy Vol.~XXI, Wiley

\bibitem[\protect\citeauthoryear{Smith}{2013}]{Smith2013} Smith N., 2013, MNRAS, 434, 102

\bibitem[\protect\citeauthoryear{Sovgut et al.}{2020}]{Sovgut2020} Sovgut, D., Krishnan, A, Trauth, A.
Fields, B.~D., 2020, in preparation

\bibitem[\protect\citeauthoryear{Stahl et al.}{2019}]{Stahl2019} Stahl B.~E., Zheng W., de Jaeger T., Filippenko A.~V., Bigley A., Blanchard K., Blanchard P.~K., et al., 2019, MNRAS, 490, 3882


\bibitem[\protect\citeauthoryear{Stephenson \& Green}{2002}]{Stephenson2002} Stephenson F.~R., Green D.~A., 2002, Historical Supernovae and Their Remnants, Oxford University Press, Oxford

\bibitem[\protect\citeauthoryear{Stephenson \& Green}{2005}]{flamsteed} Stephenson F.~R., Green D.~A., 2005, JHA, 36, 217

\bibitem[\protect\citeauthoryear{Stephenson \& Green}{2009}]{guest_stars} Stephenson F.~R., Green D.~A., 2009, JHA, 40, 31

\bibitem[\protect\citeauthoryear{Strader, et al.}{2018}]{whitepaper} Strader J., et al., 2018, arXiv, arXiv:1811.12433


\bibitem[\protect\citeauthoryear{van den Bergh}{1990}]{vandenBergh1990} van den Bergh S., 1990, AJ, 99, 843

\bibitem[\protect\citeauthoryear{van den Bergh \& Dodd}{1970}]{vandenBergh1970} van den Bergh S., Dodd W.~W., 1970, ApJ, 162, 485

\bibitem[\protect\citeauthoryear{Vink et al.}{2006}]{Vink2006} Vink J., Bleeker J., van der Heyden K., Bykov A., Bamba A., Yamazaki R., 2006, ApJL, 648, L33

\bibitem[\protect\citeauthoryear{Vink, Kaastra, \& Bleeker}{1997}]{Vink1997} Vink J., Kaastra J.~S., Bleeker J.~A.~M., 1997, A\&A, 328, 628

\bibitem[\protect\citeauthoryear{Vink}{2017}]{Vink2017} Vink, J., 2017, in Handbook of Supernovae, A.~W.~Alsabti and P.~Murdin eds, Springer, p 139

\bibitem[\protect\citeauthoryear{Wang et al.}{2020}]{Wang2020} Wang W., Siegert T., Dai Z.~G., Diehl R., Greiner J., Heger A., Krause M., et al., 2020, ApJ, 889, 169. doi:10.3847/1538-4357/ab6336

\bibitem[\protect\citeauthoryear{Wang, Fields \& Lien}{2019}]{Wang2019} Wang X., Fields B.~D., Lien A.~Y., 2019, MNRAS, 486, 2910



\end{thebibliography}




\appendix

\section{Solving Sightline Distance to a Fixed Limiting Magnitude}

\label{app:newton}

To compute and map supernova probability to an observed depth $m_{V,\rm lim}^{\rm SN}$,
for each sightline we need to find the distance $r$ that satisfies 
\begin{equation}
    m_V(r) = m_{V,\rm lim}^{\rm SN}
\end{equation}
where $m_V$ is given by eq.~(\ref{eq:appmag}).
This is the distance at which a supernova of a given absolute magnitude
has the threshold apparent magnitude in the presence of extinction.
The need to evaluate the extinction via a numerical integral
prevents an analytic solution.

This equation can be efficiently solved via root finding.
Specifically, we use Newton's method to iteratively find the root of
$f(r)=m_V(r)-m_V^{\rm lim}$,
which has derivative
\begin{eqnarray}
f^\prime(r) & = & \mu^\prime(r)+A_V^\prime(r) \\
\mu^\prime(r) & = & \frac{1}{r \ \ln 10} \\
A_V^\prime(r) & = & \kappa_V \ \rho_{\rm dust}(r) \ .
\end{eqnarray}
The method benefits from the fact that
$f^\prime$ is simply a function
evaluation,
while $f$ requires numerical
integration.
Generally we find that the solution converges after a few iterations.

\section{Monte Carlo Population}

\label{app:MC}

Using a Monte Carlo randomization technique we were able to randomly populate the Milky Way with SNE following the TRILEGAL model of stellar distribution by assigning a random position of cylindrical
radius, height, and angle drawn from our adopted rate density function for that SN type.
The angle is drawn from a uniform distribution $\theta \in [0,2\pi]$,
and the height $z$ is drawn (with both signs) from an appropriate exponential distribution in $z/h_i$
for disk component $i$.
The radial distribution is slightly more involved:  the cumulative (volume-integrated)
probability is 
\begin{equation}
    P(<R) = \frac{1}{R_i^2}\int_0^R  e^{-R^\prime/R_i} \ R \ dR 
    = 1-e^{-x}(x+1)
\end{equation}
where $x=R/R_i$.
Thus given a uniform random variable $u$,
the radius is found by solving for
\begin{equation}\label{2}
    x = \ln\frac{x+1}{1-u}
\end{equation}
Since it is impossible to isolate $x$ in this equation, we used a root finding technique to find x to a precision of $0.0001$. 

\subsection{Conversion of Coordinates}
The random generation of SNe in the Galaxy were generated with Galactocentric cylindrical coordinates $(R,\theta,z)$, but in order to calculate magnitude as viewed from the Earth these must be converted to solar-centered coordinates. 
The distance from the sun to the SN is calculated using:
\begin{equation}\label{4}
    r = \sqrt{(R \cos\theta-R_\odot)^2+(R\sin\theta)^2+(z-h_\odot)^2}
\end{equation}
The Galactic longitude for points with $R < R_\odot$ is:
\begin{equation}\label{eq:long}
    \ell =
    \left\{ 
    \begin{array}{lc}
         \arcsin \left(\frac{R\sin \theta}{r} \right) \; ,  &  R \cos \theta < R_\odot \\
         \pi - \arcsin \left(\frac{R\sin \theta}{r } \right) \; , &  R \cos \theta > R_\odot \ .
    \end{array}
    \right.
\end{equation}
The Galactic latitude can be found by:
\begin{equation}\label{6}
    b = \arcsin \left(\frac{z-z_\odot}{R} \right)
\end{equation}

\subsection{Visibility}

Using the typical absolute magnitude of a SN, the Galactocentric radius of the Sun,
and the extinction due to dust, we are able to calculate the apparent magnitude of the event at Earth. Comparing this magnitude to the minimum  apparent magnitude 
for naked-eye discovery
we are able to determine whether each randomly generated SN would be visible on Earth. Doing this calculation for thousands of SNe and then taking the ratio of visible SNe to total SNe generated, we are able to generate a predicted percentage of SNe that should be detected.

\subsection{Smoothing}
After the instances are sorted by their visibility, we then used SciPy's Gaussian kernel density estimation (KDE) function to convert the discrete events into a probability density function (PDF) for $-15^\circ<b<15^\circ$. This allows the final model for the sky distribution of supernovae to not be overfit and to stay consistent between Monte Carlo runs.

\subsection{Results:  Sky Maps}

\begin{figure}
    \centering
    \includegraphics[width=0.45\textwidth]{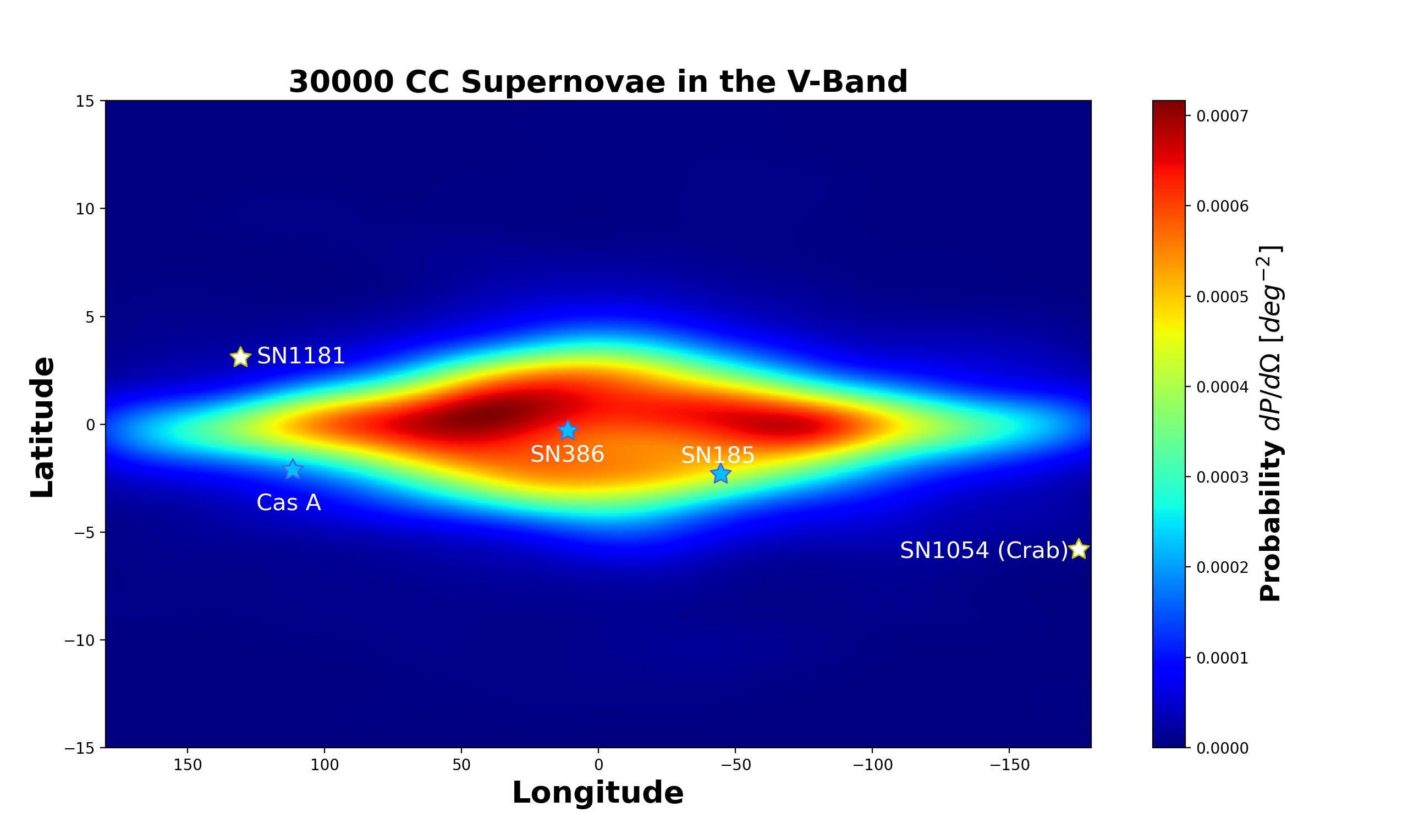}
    \caption{Probability map for naked-eye CCSNe, exactly as in Fig.~\ref{fig:cc_w_ex} but calculated using our Monte Carlo method. }
    \label{fig:CC_vis_MC}
\end{figure}

\begin{figure}
    \centering
    \includegraphics[width=0.45\textwidth]{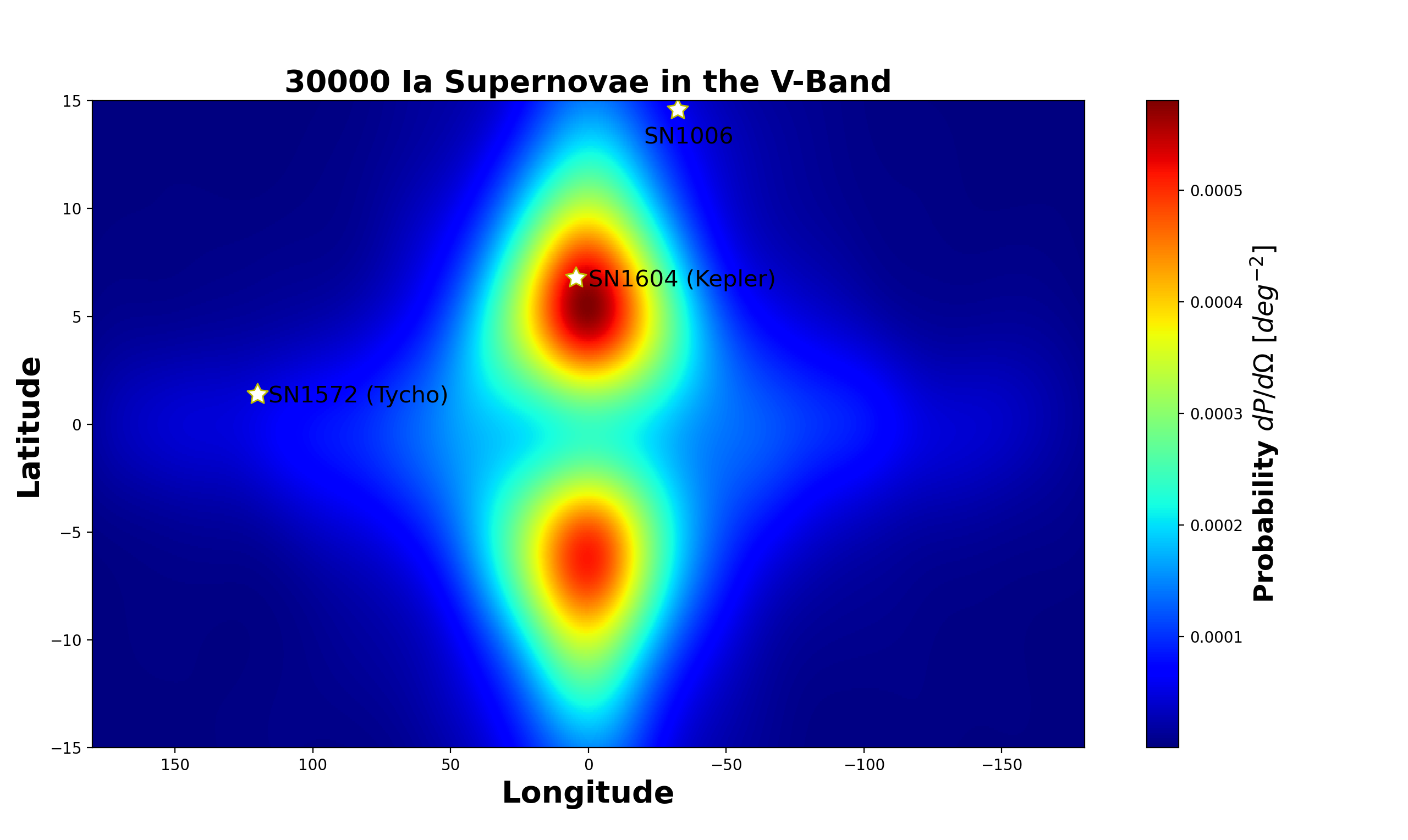}
    \caption{Monte Carlo calculation of the naked-eye SNIa sky distribution, exactly as in Fig.~\ref{fig:Ia_w_ex}. }
    \label{fig:Ia_vis_MC}
\end{figure}

We have produced Monte Carlo version of the sky probability distributions shown in the main text. In Fig.~\ref{fig:CC_vis_MC} we show
the results for naked-eye CCSN, using
30,000 simulated supernovae.
The physical situation is identical to
that shown in Fig.~\ref{fig:cc_w_ex}, to which
the Monte Carlo results should be compared.
We see good agreement generally in the overall shape and amplitude.  Note that while the integration method guarantees left-right symmetry, the Monte Carlo shows some fluctuations.  We also see that the Monte Carlo version shows more smoothing, as expected.
Finally, we see that the quantitative probability densities are in fair agreement, again with some differences due to smoothing.

Figure~\ref{fig:Ia_vis_MC} shows 
our Monte Carlo results for 30,000 
simulated SNIa, which should
be compared to Fig.~\ref{fig:Ia_w_ex}.
Here again, the qualitative and quantitative agreements are good, though the Monte Carlo smoothing is evident as in Fig.~\ref{fig:CC_vis_MC}.

\section{Alternate Supernova Distribution:  Aluminum-26 Sources}

\begin{figure}
    \centering
    \includegraphics[width=0.45\textwidth]{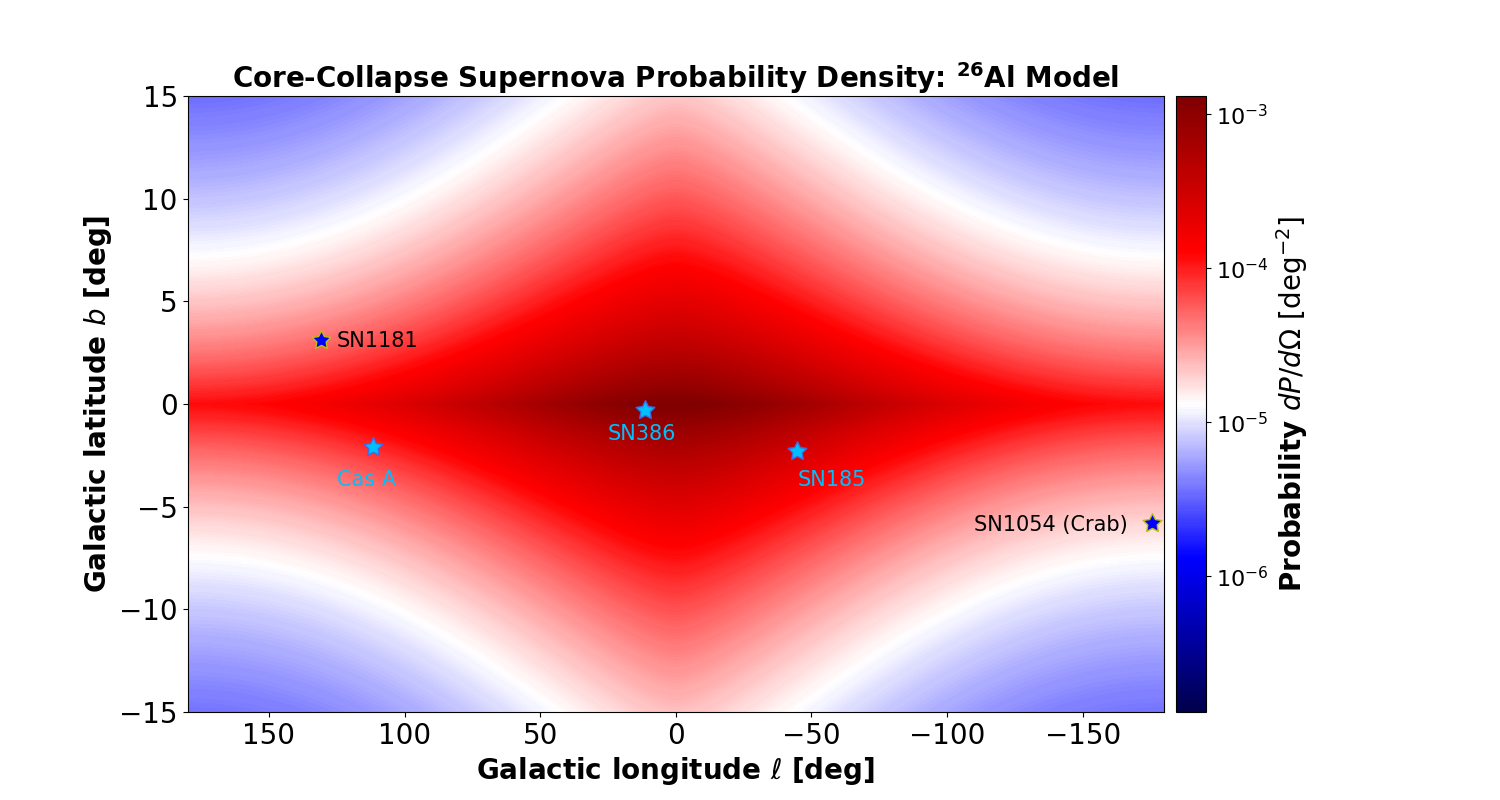}
    \caption{Intrinsic sky distribution of CCSN as in Fig.~(\ref{fig:cc_wo_ex}), but based on the
    observed \al26 sky distribution.  The large scale radius and height leads to
    a more extended map both in latitude and longitude compared to our fiducial
    result seen in Fig.~(\ref{fig:cc_wo_ex}).}
    \label{fig:CC-26Al-all}
\end{figure}

\begin{figure}
\centering
    \includegraphics[width=0.45\textwidth]{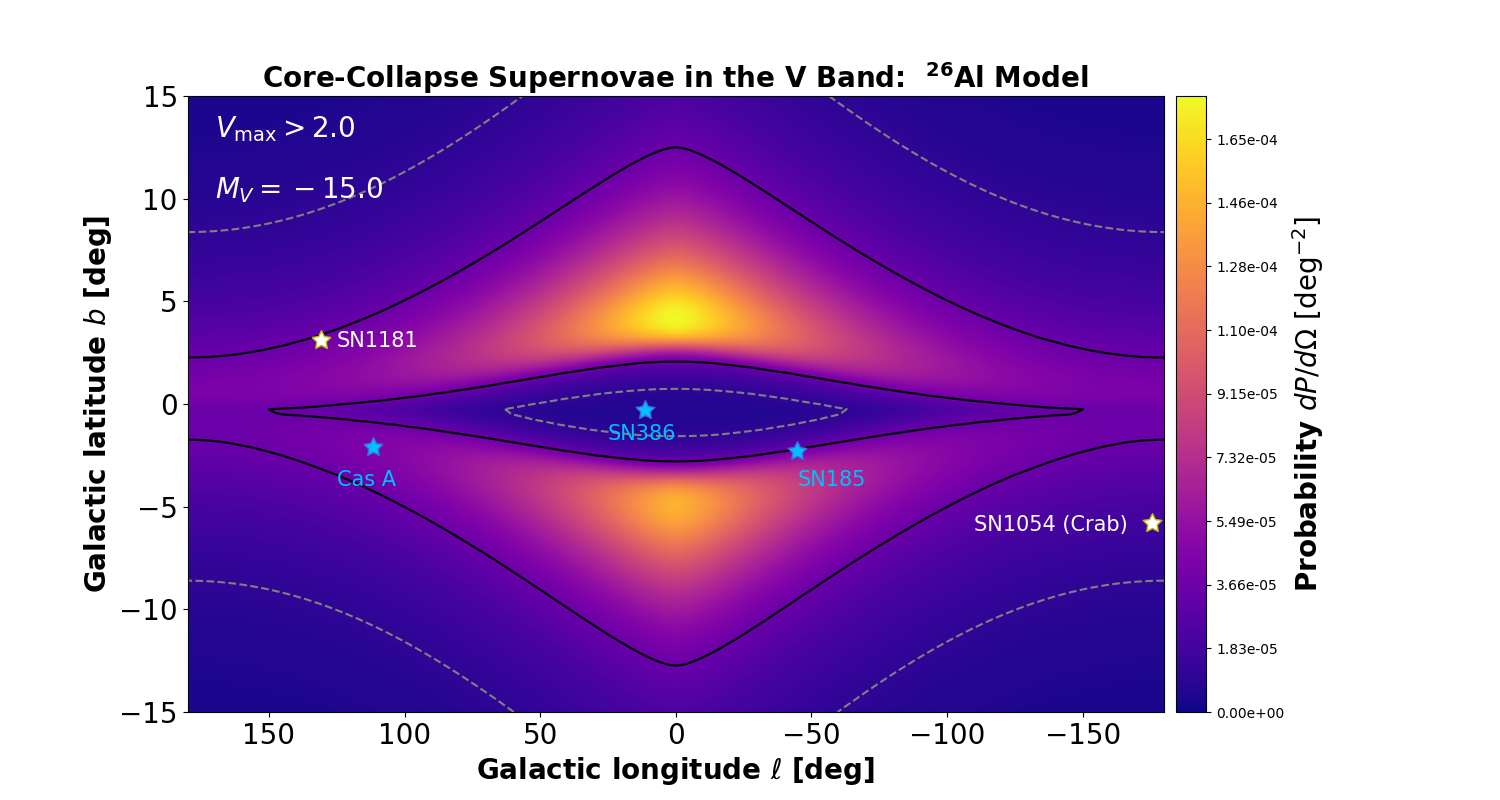}
    \caption{Naked-eye distribution of CCSN as in Fig.~\ref{fig:CC-26Al-vis},
    but based on the \al26 sky distribution.  Again we see a broader distribution in 
    latitude and longitude, and a marked deficit in the central midplane regions.}
    \label{fig:CC-26Al-vis}
\end{figure}

To illustrate the impact of the adopted supernova distribution,
here we show results for a case that represents an extreme deviation from
the our fiducial Galactic model.
As explained in \S \ref{sect:SNdist},
the supernova spatial distributions
use thin and thick disk properties
inferred from stellar counts
in the TRILEGAL model \citep{trilegal}.
The supernova remnant distribution 
implies a similar scale radius \citep{Green2015}.  But the recent
\citep{Wang2020} analysis of the \al26 gamma-ray sky
distribution
is substantially broader, with 
a much larger scale radius and 
scale height (eq.~\ref{eq:26Al}). 
Insofar as \al26 nucleosynthesis is dominated by CCSN,
this distribution would provide a direct
measure of that population.
We have therefore calculated the CCSN
sky distribution using the \al26 disk parameters to give a feel for the 
sensitivity of our results to the disk parameters, and to explore the implications of the \al26 results.
We retain the dust distribution 
in eq.~(\ref{eq:dustdist}),
so that the dust is much more
spatially compact than the supernovae
themselves.

Figure \ref{fig:CC-26Al-all} shows the intrinsic sky distribution of
supernovae that follow the \al26 
distribution.  This is to be compared to
the results from our fiducial case 
in Fig.~\ref{fig:cc_wo_ex}; recall
that these plots use a logarthimic scale
for the probability density indicated
by the color range.
We see that the \al26 distribution has
a wider span in Galactic longitude, and extends to higher latitudes.  
These are expected given the larger scale radius and height inferred from
\al26 .
While the Galactic center still has the maximum probability in the \al26 case, the peak
value is lower than in the fiducial case.
This reflects more diffuse nature of the \al26 map, and the requirement that the distribution is normalized 
to $\int dP/d\Omega \ d\Omega = 1$.

Figure~\ref{fig:CC-26Al-vis} shows the sky distribution of naked-eye CCSN that
follow the \al26 disk model.
This is to be compared to the fiducial results in Fig.~\ref{fig:cc_w_ex}.
The differences are dramatic.
The peak regions are in the inner quadrants but at $|b| \sim 5^\circ$ off
of the plane.
Also notable is a {\em deep minimum}
in probability at and around the Galactic midplane.  Both features 
are more
extreme version of effects seen in our
fiducial SNIa distribution in (Fig.~\ref{fig:Ia_w_ex},
and trace back to similar causes.
These reflect the more dilute supernova distribution
in the \al26 picture:  compared to
the fiducial case, there are fewer supernovae along the midplane compared
and more at larger heights.  Thus, 
the more compact nature of the
dust distribution leads to strong
extinction at low latitudes
but leaves the high latitudes unaffected.

Turning to the historical supernova locations in \ref{fig:CC-26Al-vis},
we see that even in this model with supernovae extending to large radii and heights, still none of the supernovae lie in the peak regions.  That said,
the Crab, SN 1181, and Cas A lie in
regions of higher probability than in Fig.~\ref{fig:cc_w_ex},
and the Crab is now inside the 95\% contour.  This reflects the high latitudes of these events.  On the other hand, the low-latitude locations of SN185 and 386 are {\em disfavored} in this model, with SN 396 now outside of the 95\% contour.


\bsp	
\label{lastpage}
\end{document}